\documentclass[12pt]{article}

\usepackage[english]{babel}
\usepackage[utf8x]{inputenc}
\usepackage{amsmath}
\usepackage{graphicx}
\usepackage{etoolbox}
\usepackage{changepage}
\usepackage{titlesec}
\usepackage[parfill]{parskip}
\usepackage[margin=1in]{geometry}
\usepackage{times}
\usepackage{float}
\usepackage[numbers,square]{natbib}

\usepackage{multirow}
\usepackage{booktabs}
\usepackage{wrapfig}
\usepackage{makecell}
\usepackage{graphicx}
\usepackage{subcaption}
\PassOptionsToPackage{hyphens}{url}\usepackage{hyperref}

\titleformat*{\section}{\normalsize\bfseries} 

\title{\large \textbf{Assessing Student Adoption of Generative Artificial Intelligence across Engineering Education from 2023 to 2024}} 

\author{Jesan Ahammed Ovi, Gabe Fierro, and C. Estelle Smith}

\date{} 

\makeatletter 
\patchcmd{\@maketitle}{\begin{center}}{\begin{adjustwidth}{0.5in}{0.5in}\begin{center}}{}{}
\patchcmd{\@maketitle}{\end{center}}{\end{center}\end{adjustwidth}}{}{}
\makeatother

\newcommand{\participants}{
\begin{table}[]
\footnotesize
\begin{tabular}{|r|lllll|}
\hline
\multicolumn{1}{|l|}{} &
  \multicolumn{5}{c|}{\textbf{2023 SURVEY ($n_1=601$): Undergraduate (474, 78.9\%); Graduate (127, 21.1\%)}} \\ \hline
\textbf{Dept. Cluster} &
  \multicolumn{1}{l|}{\makecell{\textbf{Mech-Civil} \\ 152 (25.3\%)}} & 
  \multicolumn{1}{l|}{\makecell{\textbf{CS-EE-AMS} \\ 200 (33.3\%)}} & 
  \multicolumn{1}{l|}{\makecell{\textbf{Met-Geo-Pet} \\ 108 (18\%)}} & 
  \multicolumn{1}{l|}{\makecell{\textbf{Phys-Chem} \\ 133 (22.1\%)}} & 
  \makecell{\textbf{Society} \\ 8 (1.3\%)} \\ \hline
\textbf{\# Years} &
  \multicolumn{1}{l|}{\makecell{0-1 Yrs \\ 166 (27.6\%)}} &
  \multicolumn{1}{l|}{\makecell{1-2 Yrs \\ 164 (27.3\%)}} &
  \multicolumn{1}{l|}{\makecell{2-3 Yrs \\ 122 (20.3\%)}} &
  \multicolumn{1}{l|}{\makecell{3-4 Yrs \\ 97 (16.1\%)}} &
  \makecell{4+ Yrs \\ 52 (8.7\%)} \\ \hline \hline
\textbf{} &
  \multicolumn{5}{c|}{\textbf{2024 SURVEY ($n_2=862$): Undergraduate (647, 75.1\%); Graduate (215, 24.9\%)}} \\ \hline 
\textbf{Dept. Cluster} &
  \multicolumn{1}{l|}{\makecell{\textbf{Mech-Civil} \\ 315 (36.5\%)}} & 
  \multicolumn{1}{l|}{\makecell{\textbf{CS-EE-AMS} \\ 233 (27\%)}} & 
  \multicolumn{1}{l|}{\makecell{\textbf{Met-Geo-Pet} \\ 113 (13.1\%)}} & 
  \multicolumn{1}{l|}{\makecell{\textbf{Phys-Chem} \\ 167 (19.4\%)}} & 
  \makecell{\textbf{Society} \\ 34 (3.9\%)} \\ \hline 
\textbf{\# Years} &
  \multicolumn{1}{l|}{\makecell{0-1 Yrs \\ 323 (37.5\%)}} &
  \multicolumn{1}{l|}{\makecell{1-2 Yrs \\ 179 (20.8\%)}} &
  \multicolumn{1}{l|}{\makecell{2-3 Yrs \\ 163 (18.9\%)}} &
  \multicolumn{1}{l|}{\makecell{3-4 Yrs \\ 124 (14.4\%)}} &
  \makecell{4+ Yrs \\ 73 (8.5\%)} \\ \hline
\textbf{Race} &
  \multicolumn{1}{l|}{\makecell{White \\ 667 (77.4\%)}} &
  \multicolumn{1}{l|}{\makecell{Asian/Pacific Islander \\ 109 (12.7\%)}} &
  \multicolumn{1}{l|}{\makecell{Hispanic/Latinx \\ 93 (10.8\%)}} &
  \multicolumn{1}{l|}{\makecell{Black/African American \\ 25 (2.9\%)}} &
  \makecell{Other \\ 57 (6.6\%)} \\ \hline
\textbf{Gender} &
  \multicolumn{1}{l|}{\makecell{Male \\ 495 (57.4\%)}} &
  \multicolumn{1}{l|}{\makecell{Female \\ 283 (32.8\%)}} &
  \multicolumn{1}{l|}{\makecell{Non-binary \\ 42 (4.9\%)}} &
  \multicolumn{1}{l|}{\makecell{Preferred no reply \\ 42 (4.9\%)}} &
  \multicolumn{1}{c|}{-} \\ \hline
\textbf{SES} &
  \multicolumn{1}{l|}{\makecell{Low (1-3) \\ 54 (6.4\%)}} &
  \multicolumn{1}{l|}{\makecell{Moderate (4-7) \\ 593 (69.6\%)}} &
  \multicolumn{1}{l|}{\makecell{High (8-10) \\ 204 (24\%)}} &
  \multicolumn{1}{c|}{-} &
  \multicolumn{1}{c|}{-} \\ \hline
\end{tabular}
\caption{\textit{Participant Demographics. Socioeconomic Status (SES) reported using MacArthur Scale Ratings. Row for race sums to $>$100\% because participants could select $>$1 race/ethnicity.}}
\label{table:respondents}
\end{table}
}

\newcommand{\deptclusters}{
        \begin{table}[]
        \footnotesize
        \centering
        \renewcommand{\arraystretch}{1.2} 
        \setlength{\tabcolsep}{4pt} 
        \begin{tabular}{|p{0.3\textwidth}|p{0.6\textwidth}|}
            \hline
            \textbf{Name of Cluster} & \textbf{Departments Included} \\ \hline
            \textbf{Mech-Civil} 
            & Mechanical Eng., Civil \& Environmental Eng. \\ \hline
            \textbf{CS-EE-AMS} 
            & Computer Science, Electrical Eng., Applied Math \& Stats \\ \hline
            \textbf{Met-Geo-Pet} 
            & Metallurgical Eng., Geology, Geophysics, Mining, Petroleum Eng. \\ \hline
            \textbf{Phys-Chem} 
            & Physics, Chemistry, Chemical \& Biological Eng. \\ \hline
            \textbf{Society} 
            & Engineering Design \& Society, Economics \& Business, Humanities \\ \hline
        \end{tabular}
        \caption{\textit{Department Clusters Used in Analysis of Student Adoption of GenAI.}}
        \label{table:department_clusters}
        \end{table}
}

\newcommand{\adoptionrate}{
\begin{wrapfigure}{l}{0.6\textwidth}
    \centering
    \includegraphics[width=0.58\textwidth]{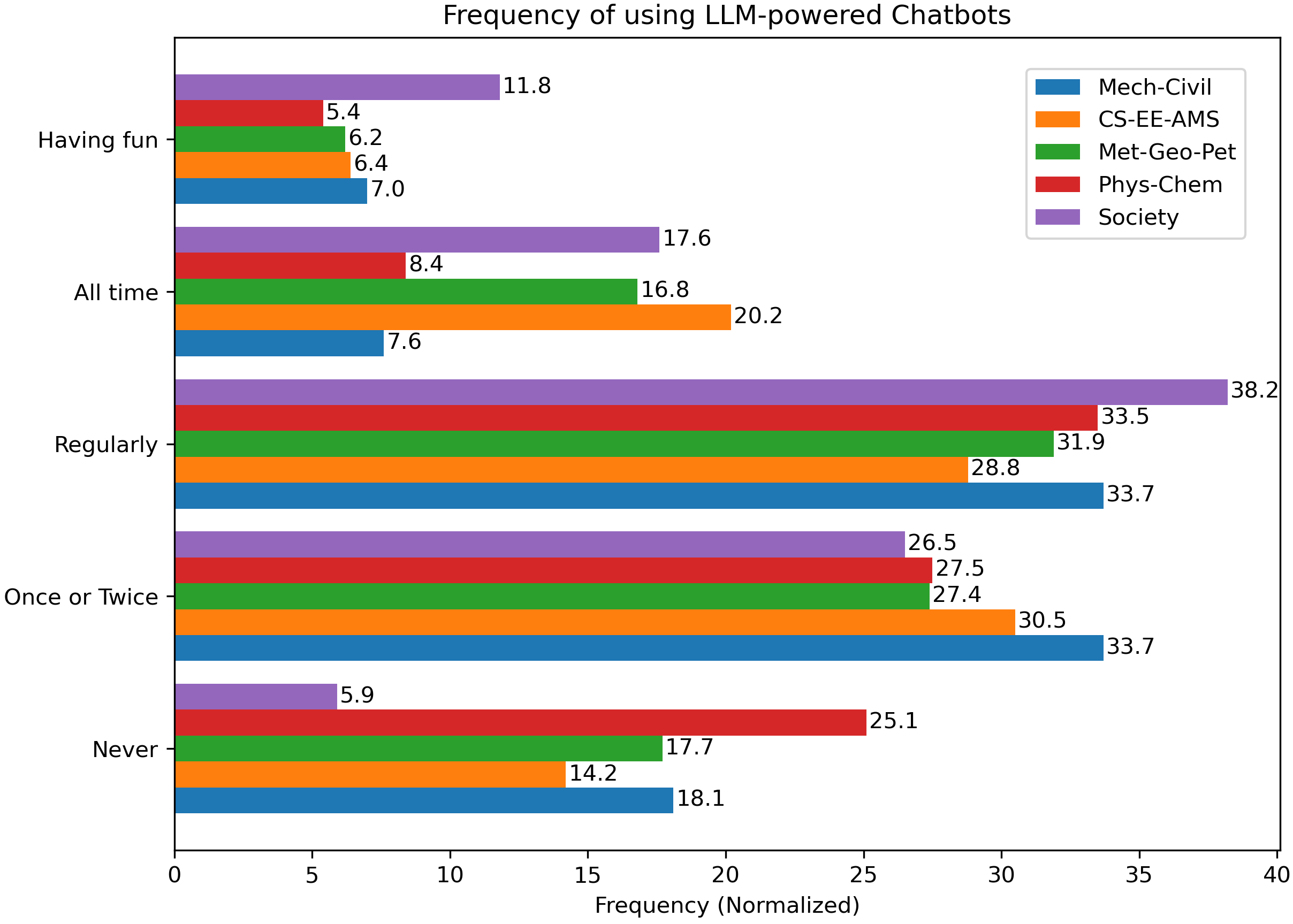}
    \caption{\textit{Rate of Using LLM-Chatbots in 2024. X-axis: Frequency normalized by \# of responses/dept. cluster.}}
    \label{fig:frequencyusaecase}
    \vspace{-10pt}
\end{wrapfigure}
}

\newcommand{\usecases}{
\begin{figure}
    \centering
    \begin{subfigure}[b]{\textwidth}
        \centering
        \includegraphics[width=\textwidth]{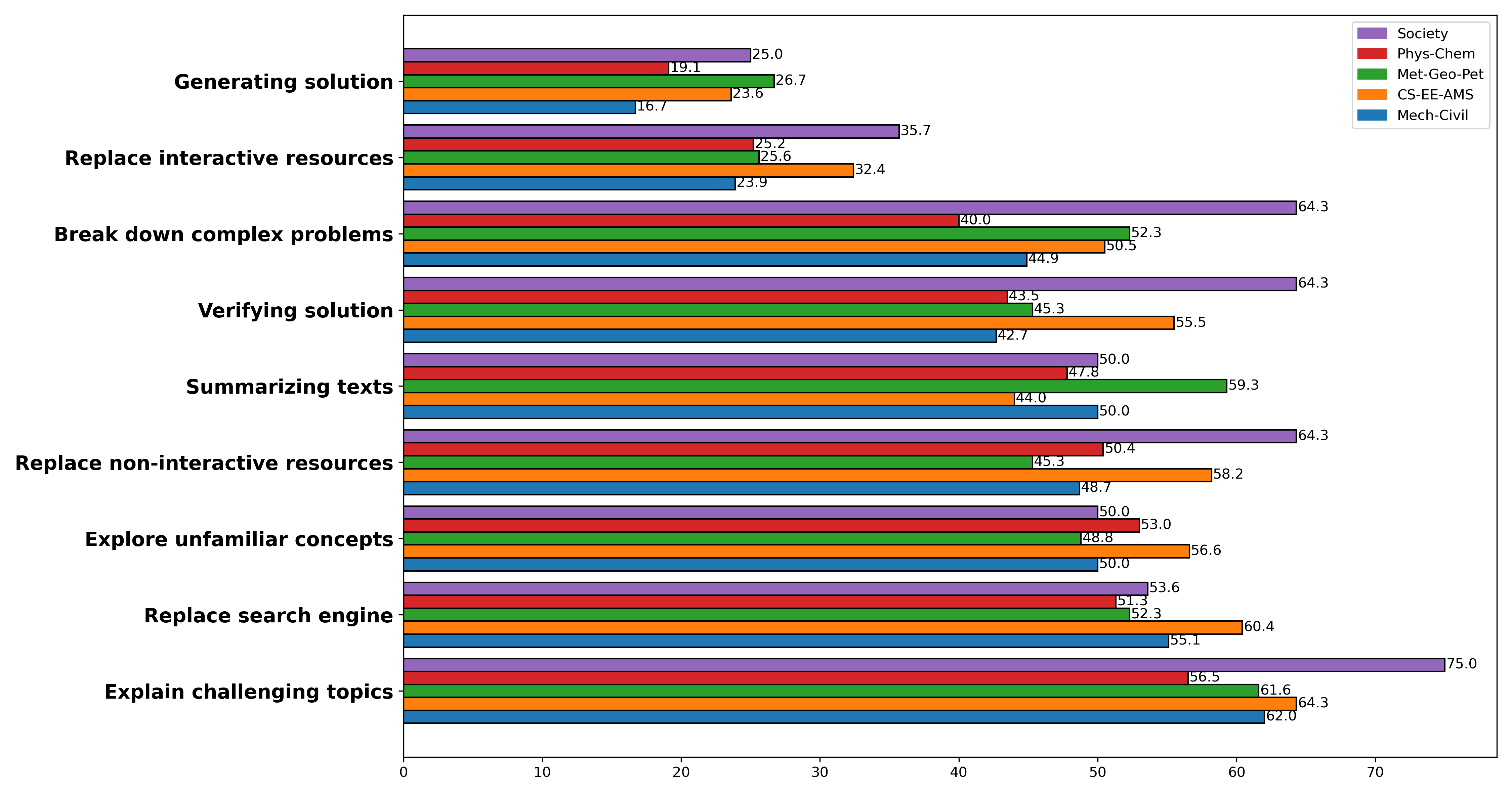}
        \caption{\textit{Learning Use Cases.}}
        \label{fig:learning}
    \end{subfigure}
    \begin{subfigure}[b]{\textwidth}
        \centering
        \includegraphics[width=\textwidth]{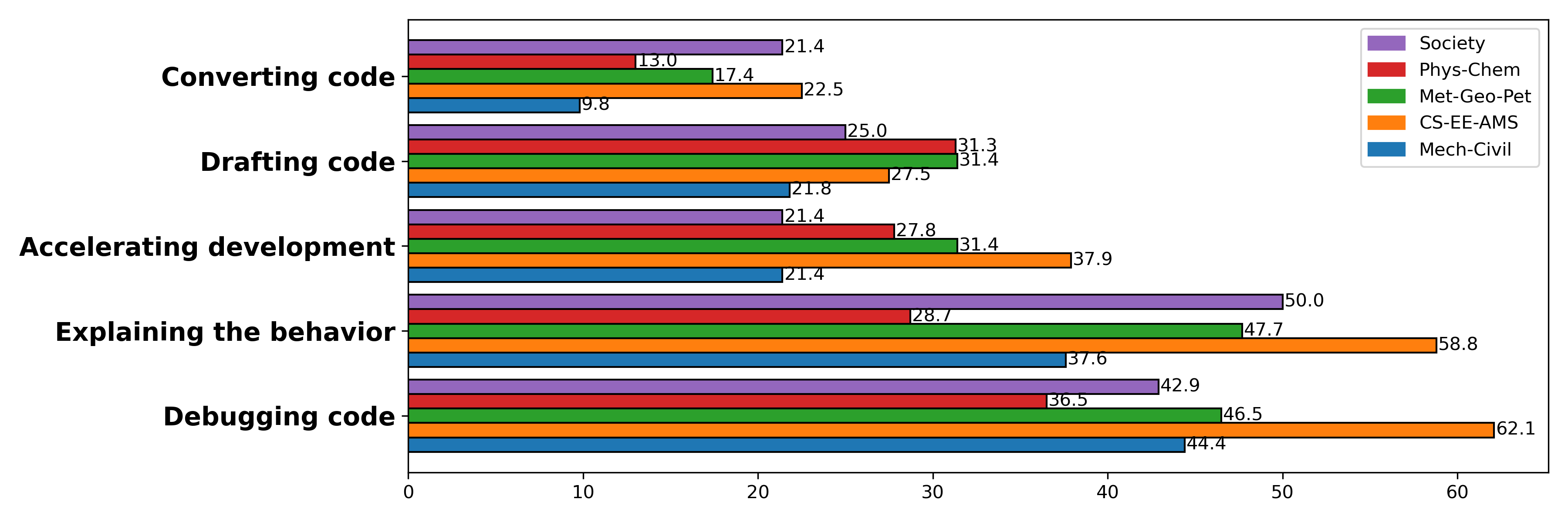} 
        \caption{\textit{Coding Use Cases.}}
        \label{fig:coding}
    \end{subfigure}
    \begin{subfigure}[b]{\textwidth}
        \centering
        \includegraphics[width=\textwidth]{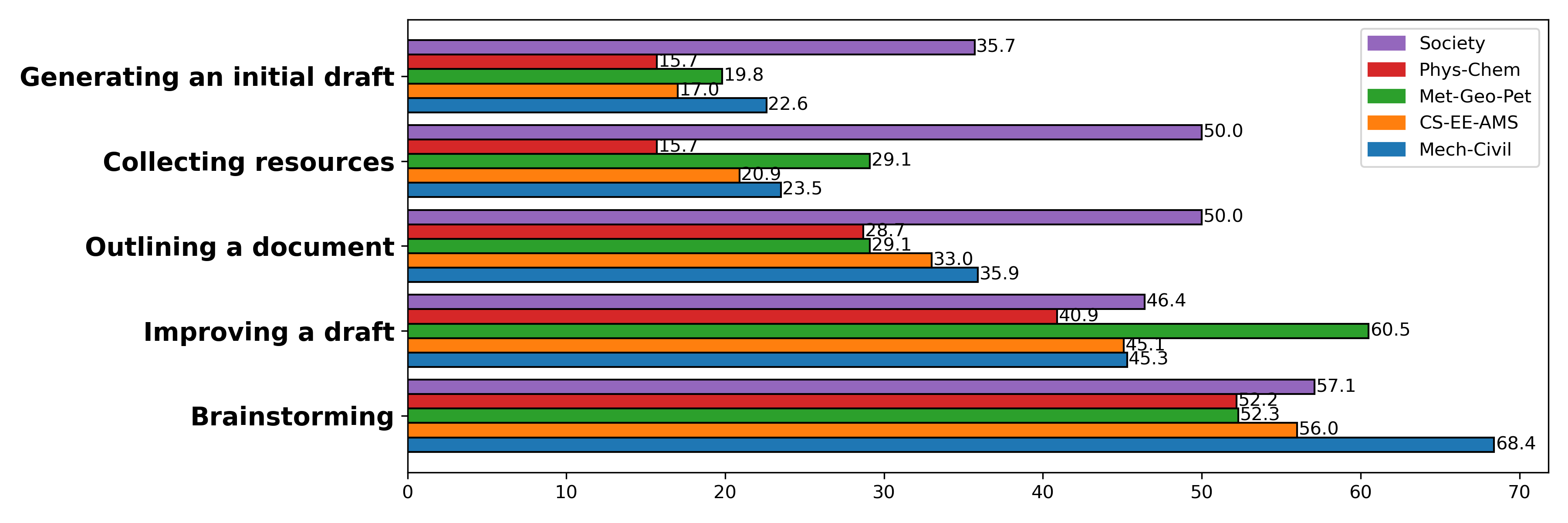}
        \caption{\textit{Writing Use Cases.}}
        \label{fig:writing}
    \end{subfigure}
    \caption{\textit{Use Cases for LLM-Chatbots in Engineering Education in 2024 ($n=651$). Y-axes: Use case labels. X-axes: Frequencies normalized by \# of respondents per department cluster.}}
    \label{fig:usecases}
\end{figure}
}

\newcommand{\scientificfield}{
\begin{figure}[t]
    \centering
    \begin{subfigure}[b]{0.49\linewidth}
        \centering
        \includegraphics[width=\linewidth]{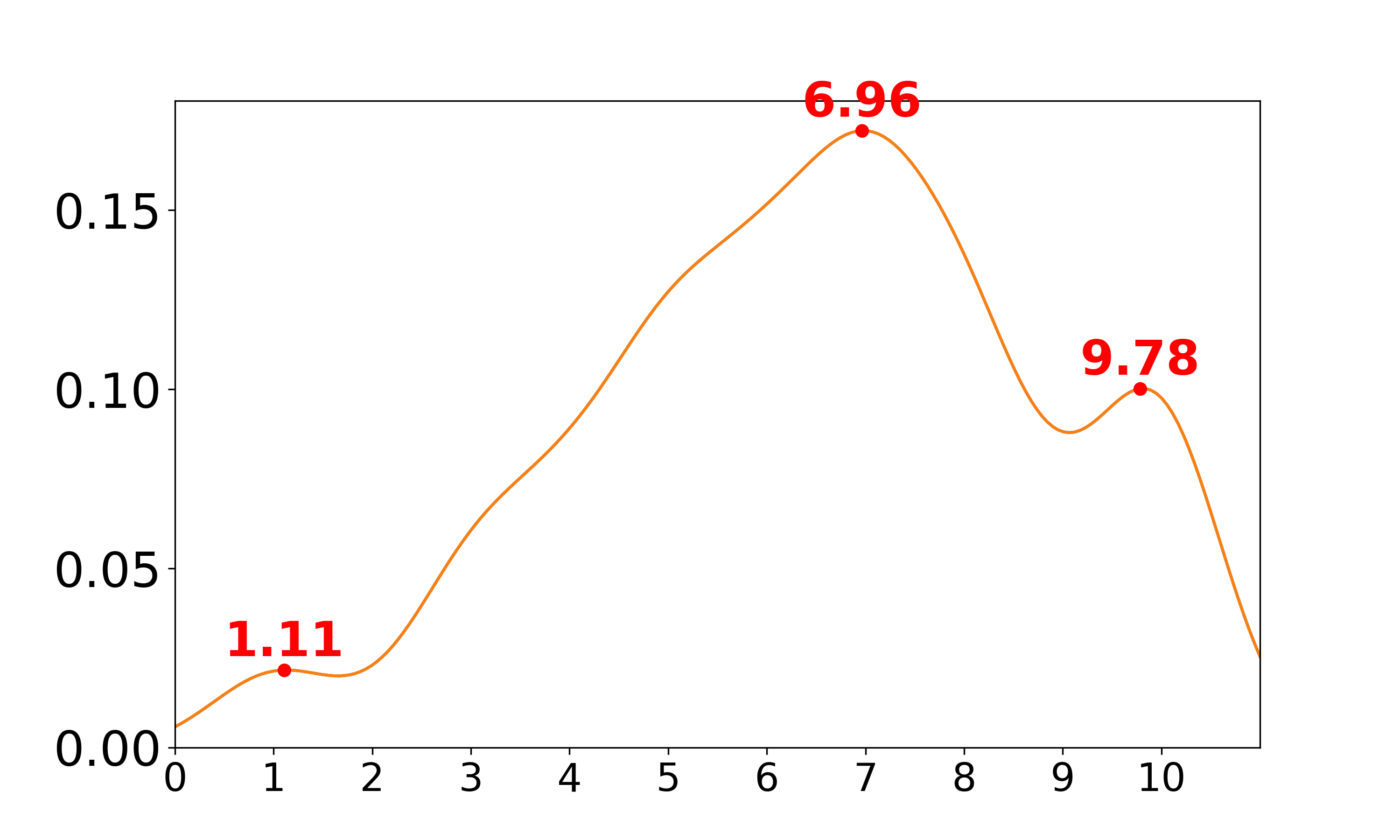} 
        \caption{\textit{Student Ratings in 2023 ($n=601$).}}
        \label{fig:fieldbenefit2023}
    \end{subfigure}
    \hfill
    \begin{subfigure}[b]{0.49\linewidth}
        \centering
        \includegraphics[width=\linewidth]{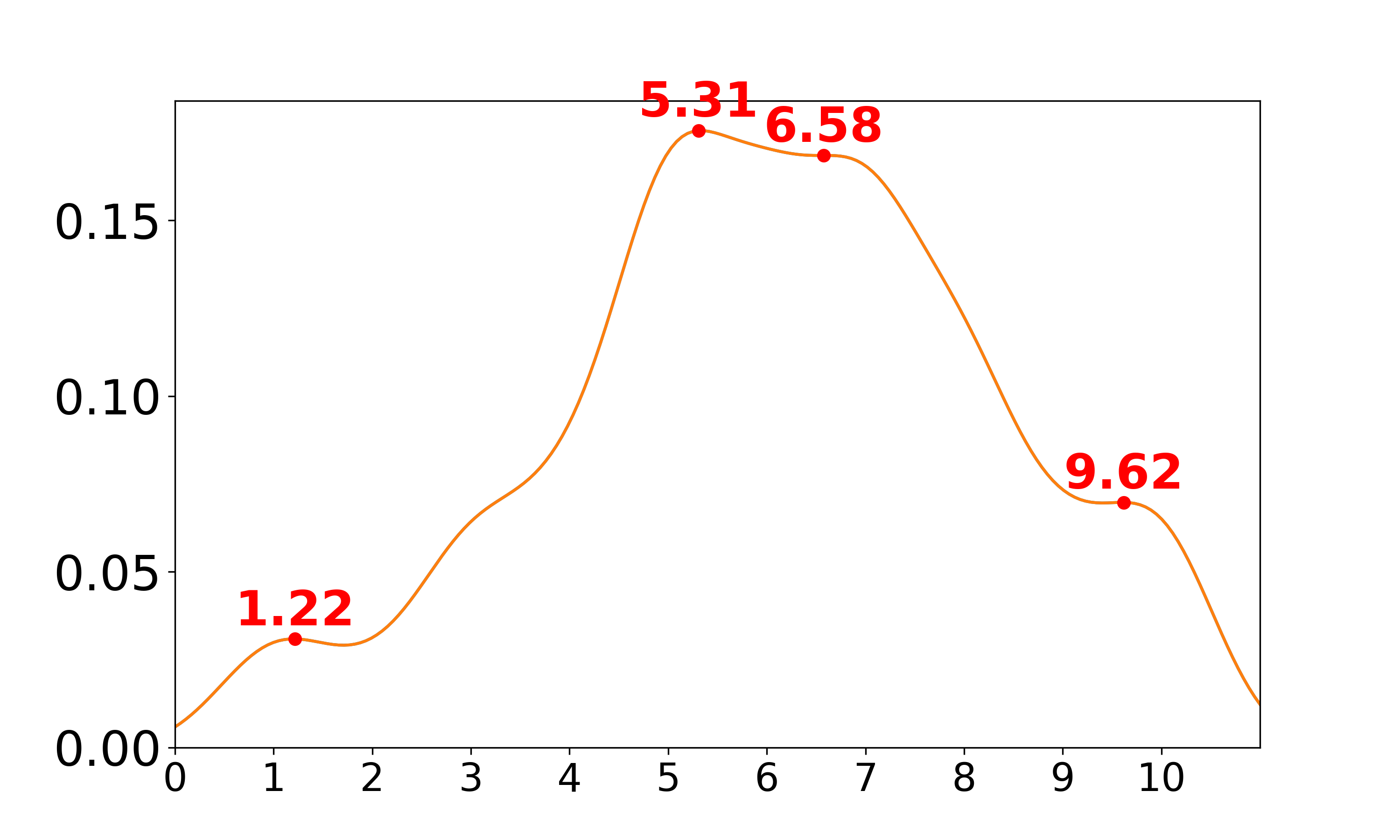} 
        \caption{\textit{Student Ratings in 2024 ($n=861$).}}
        \label{fig:fieldbenefit2024}
    \end{subfigure}
    \caption{\textit{KDE Plots of Perceived Benefits \textit{v.s.} Harms to `` Your Scientific Field or Major.'' X-axes: Likert ratings from 1 (extremely harmful) to 10 (extremely beneficial). Y-axes: Densities.}}
    \label{fig:fieldbenefits}
\end{figure}
}

\newcommand{\ratings}{
    \begin{figure}[t]
        \centering
        \begin{subfigure}[b]{0.3\textwidth}
            \includegraphics[width=\textwidth]{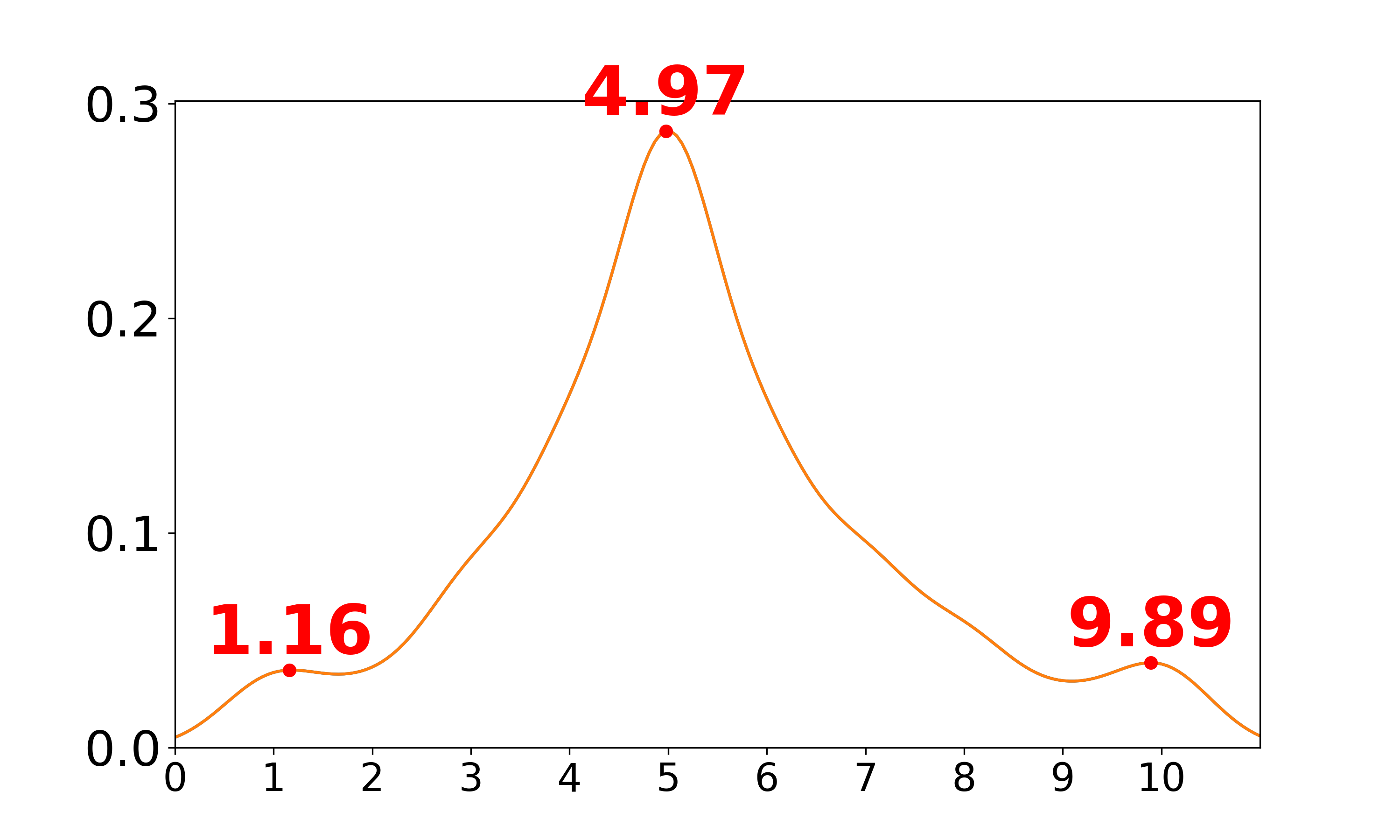} 
            \caption{\textit{``Your Job Placement.''}}
            \label{fig:jobplacement}
        \end{subfigure}
        \hfill
        \begin{subfigure}[b]{0.3\textwidth}
            \includegraphics[width=\textwidth]{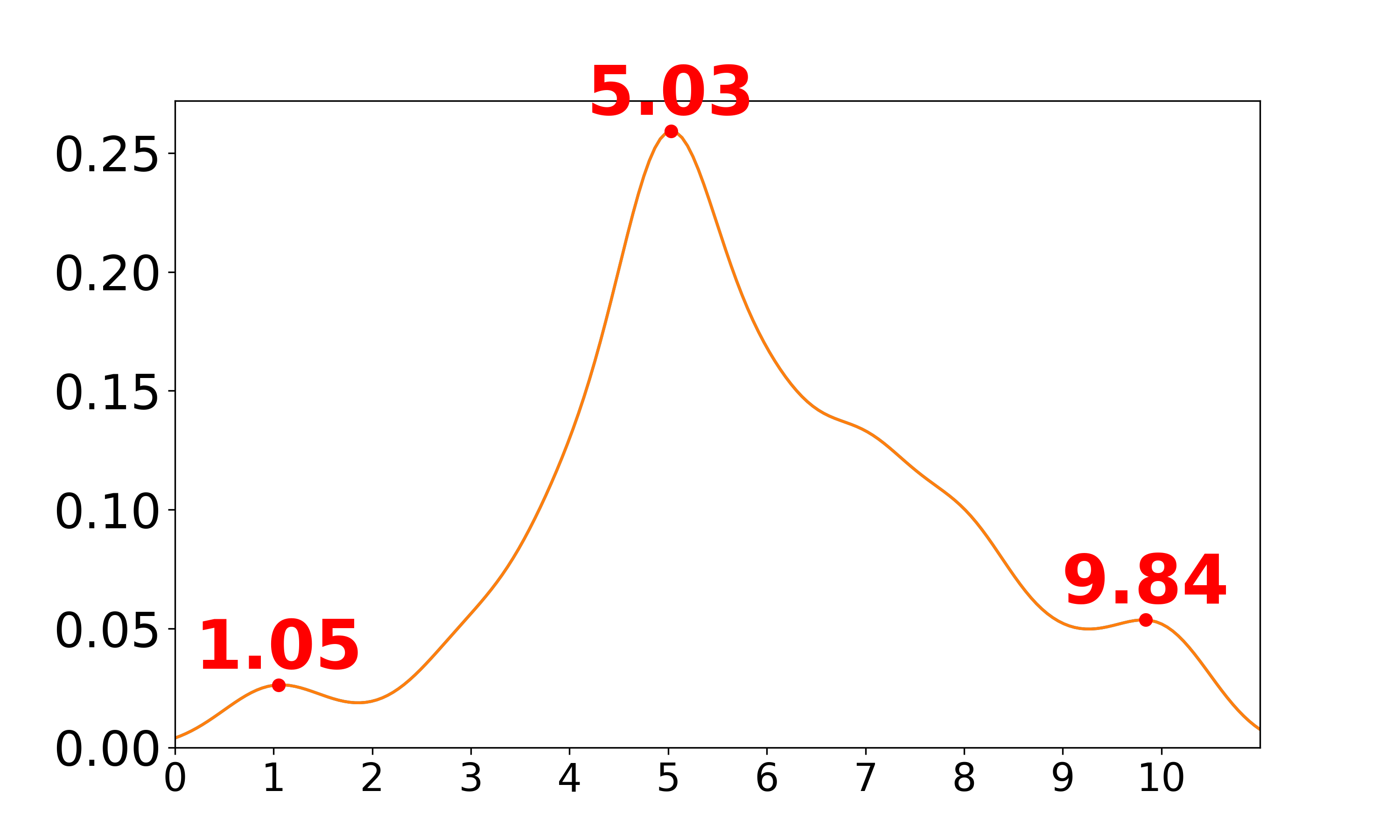} 
            \caption{\textit{``Your Intended Job Itself.''}}
            \label{fig:intendedjob}
        \end{subfigure}
        \hfill
        \begin{subfigure}[b]{0.3\textwidth}
            \includegraphics[width=\textwidth]{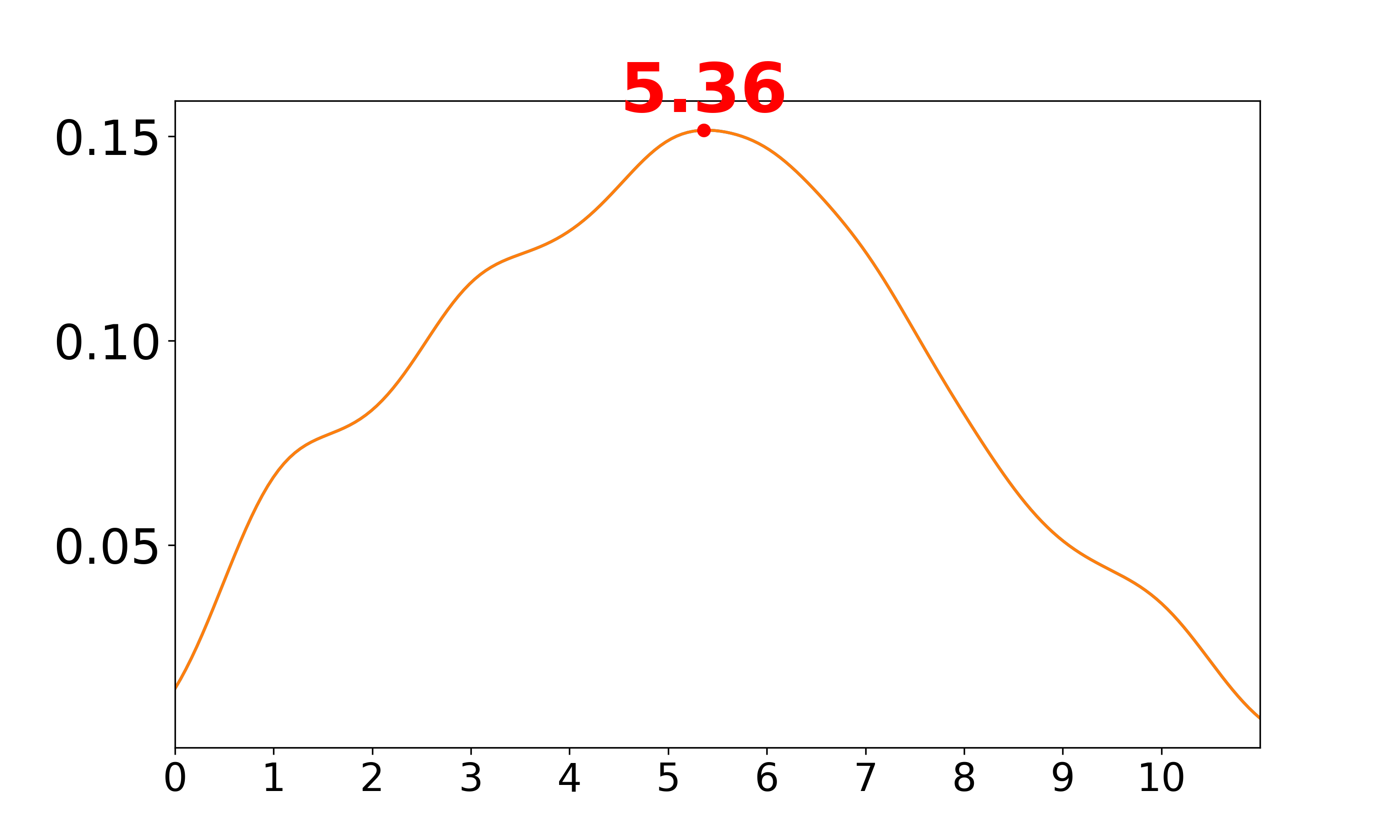} 
            \caption{\textit{``Society at Large.''}}
            \label{fig:societyatlarge}
        \end{subfigure}
               \begin{subfigure}[b]{0.3\textwidth}
            \centering
            \includegraphics[width=\textwidth]{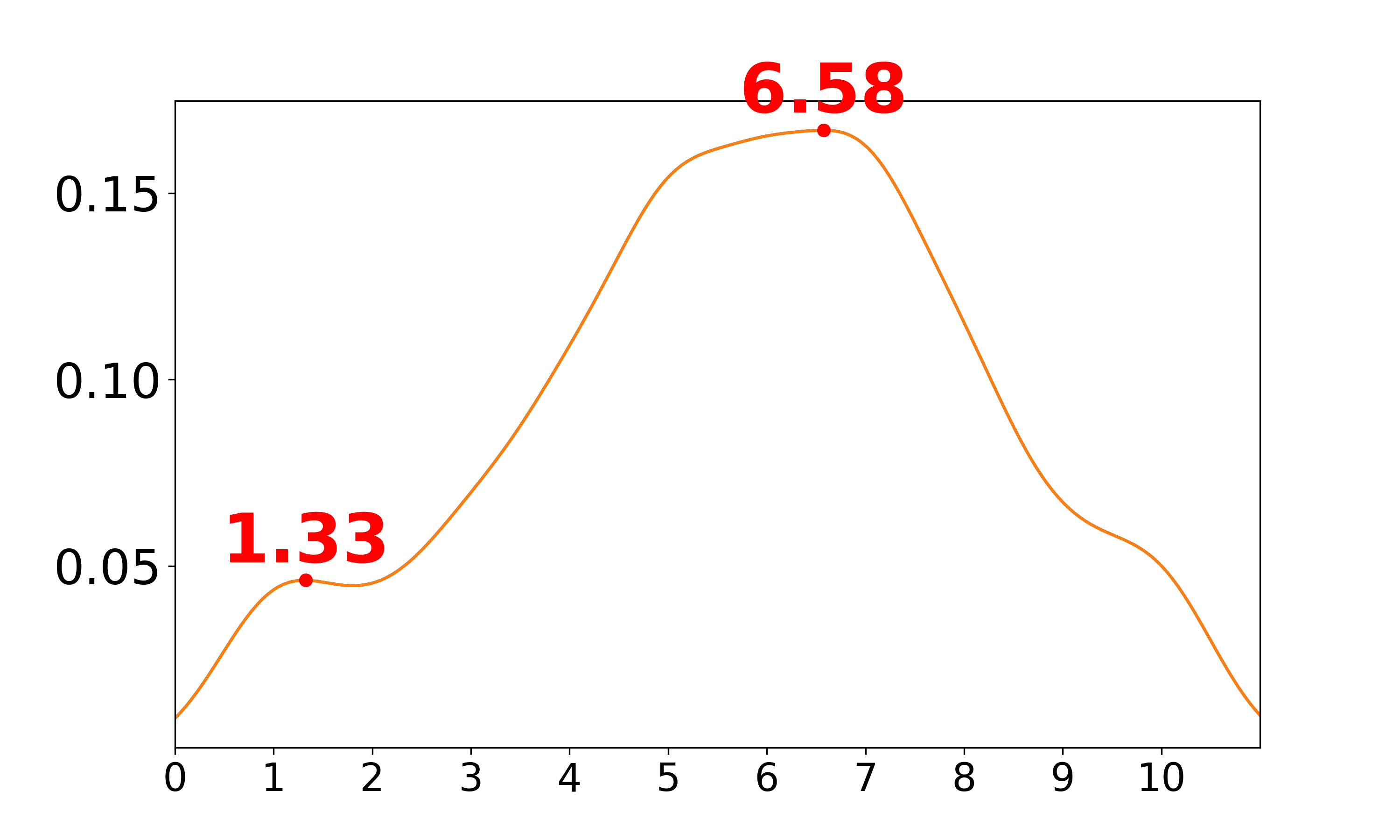} 
            \caption{\textit{``Your Learning During Higher Education.''}}
            \label{fig:yourlearning}
        \end{subfigure}
        \hfill
        \begin{subfigure}[b]{0.3\textwidth}
            \centering
            \includegraphics[width=\textwidth]{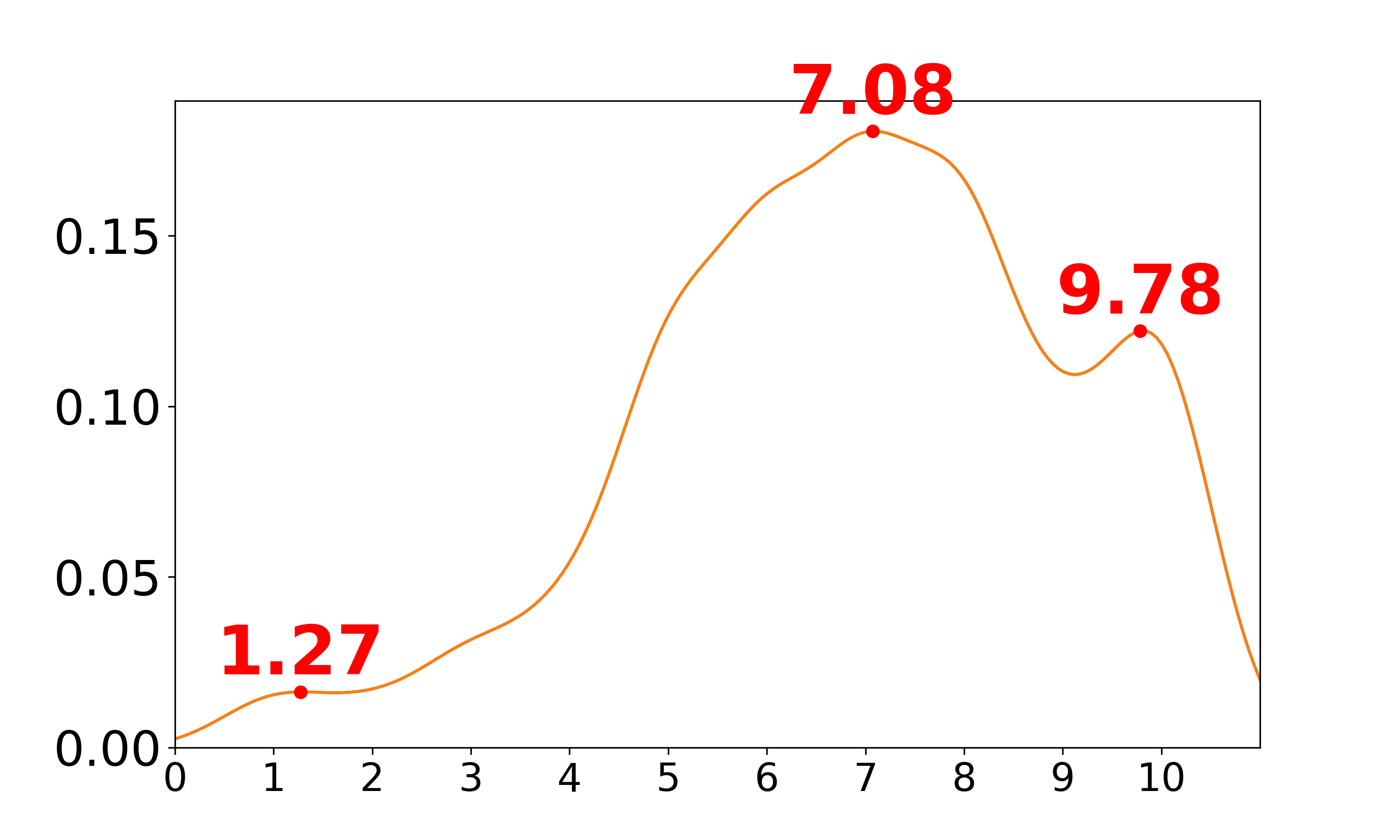} 
            \caption{\textit{``Your Efficiency at Completing Tasks.''}}
            \label{fig:yourefficiency}
        \end{subfigure}
        \hfill
        \begin{subfigure}[b]{0.3\textwidth}
            \centering
            \includegraphics[width=\textwidth]{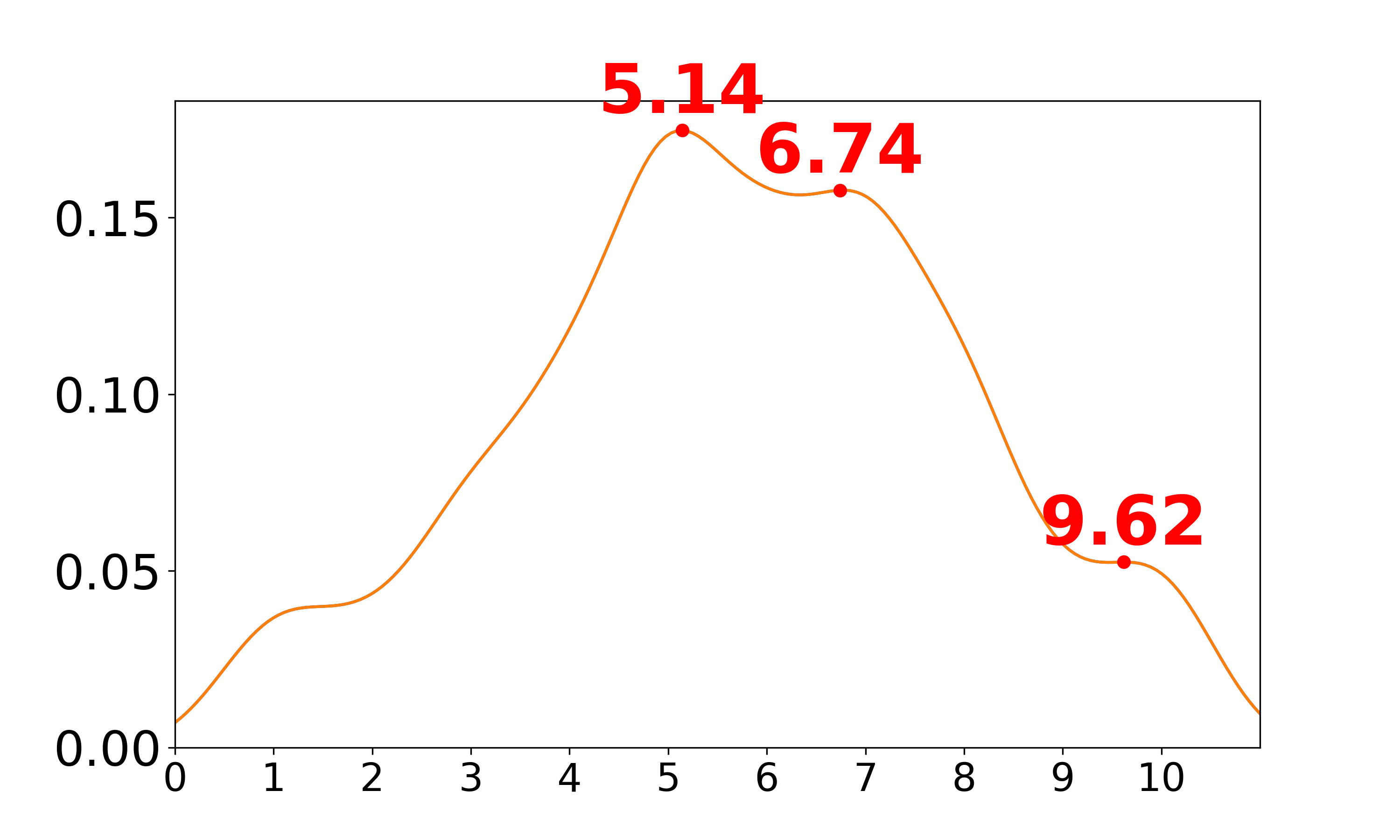} 
            \caption{\textit{``Quality of the Tasks You Complete.''}}
            \label{fig:taskquality}
        \end{subfigure}
        \caption{\textit{KDE Plots of Perceived Benefits \textit{v.s.} Harms of GenAI in 2024 ($n=862$). X-axes:  Likert ratings from 1 (extremely harmful) to 10 (extremely beneficial). Y-axes: Densities.}}
        \label{fig:perception1}
    \end{figure}
}

\newcommand{\motivations}{
\begin{wrapfigure}{r}{0.6\textwidth}
    \centering
    \includegraphics[width=0.58\textwidth]{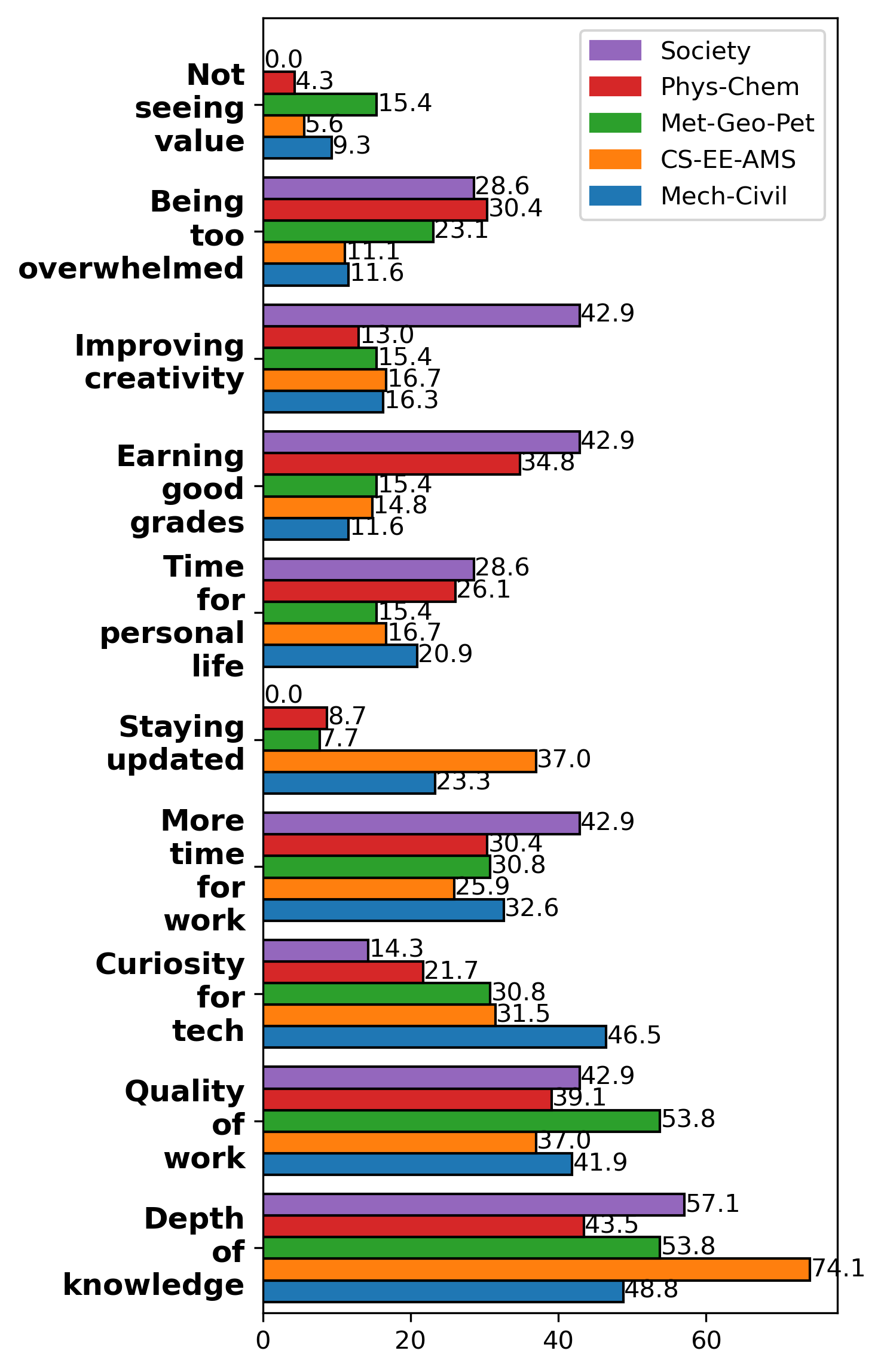}
    \caption{\textit{Students' Top Motivations for Using LLM-Chatbots in 2024 ($n=142$). X-axis: Frequencies normalized by \# of respondents per dept. cluster.}}
    \label{fig:motivation}
\end{wrapfigure}
}

\newcommand{\studentethics}{
\begin{wrapfigure}{r}{0.58\textwidth}
    \vspace{-10pt}
    \centering
    \includegraphics[width=0.57\textwidth]{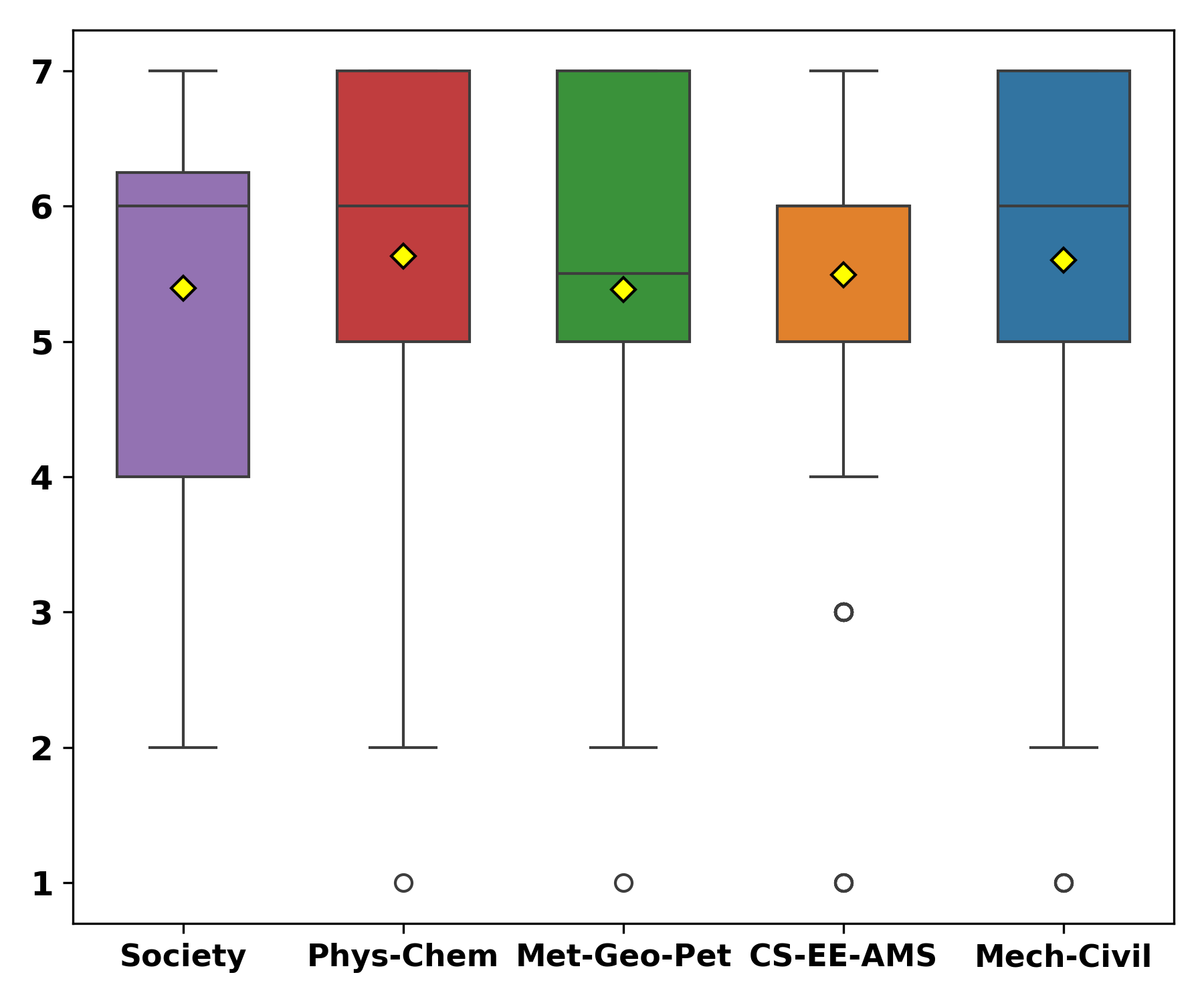}
    \caption{\textit{Boxplots of Student Ratings of Ethical Use of LLMs in 2024 ($n=651$). Y-axis: Likert ratings from 1 (completely unethical) to 7 (completely ethical). X-axis: Department Clusters.}}
    \label{fig:ethical}
    \vspace{-10pt}
\end{wrapfigure}
}

\newcommand{\concerns}{
\begin{wrapfigure}{l}{0.57\textwidth}
    \centering
    \vspace{-14pt}
    \includegraphics[width=0.56\textwidth]{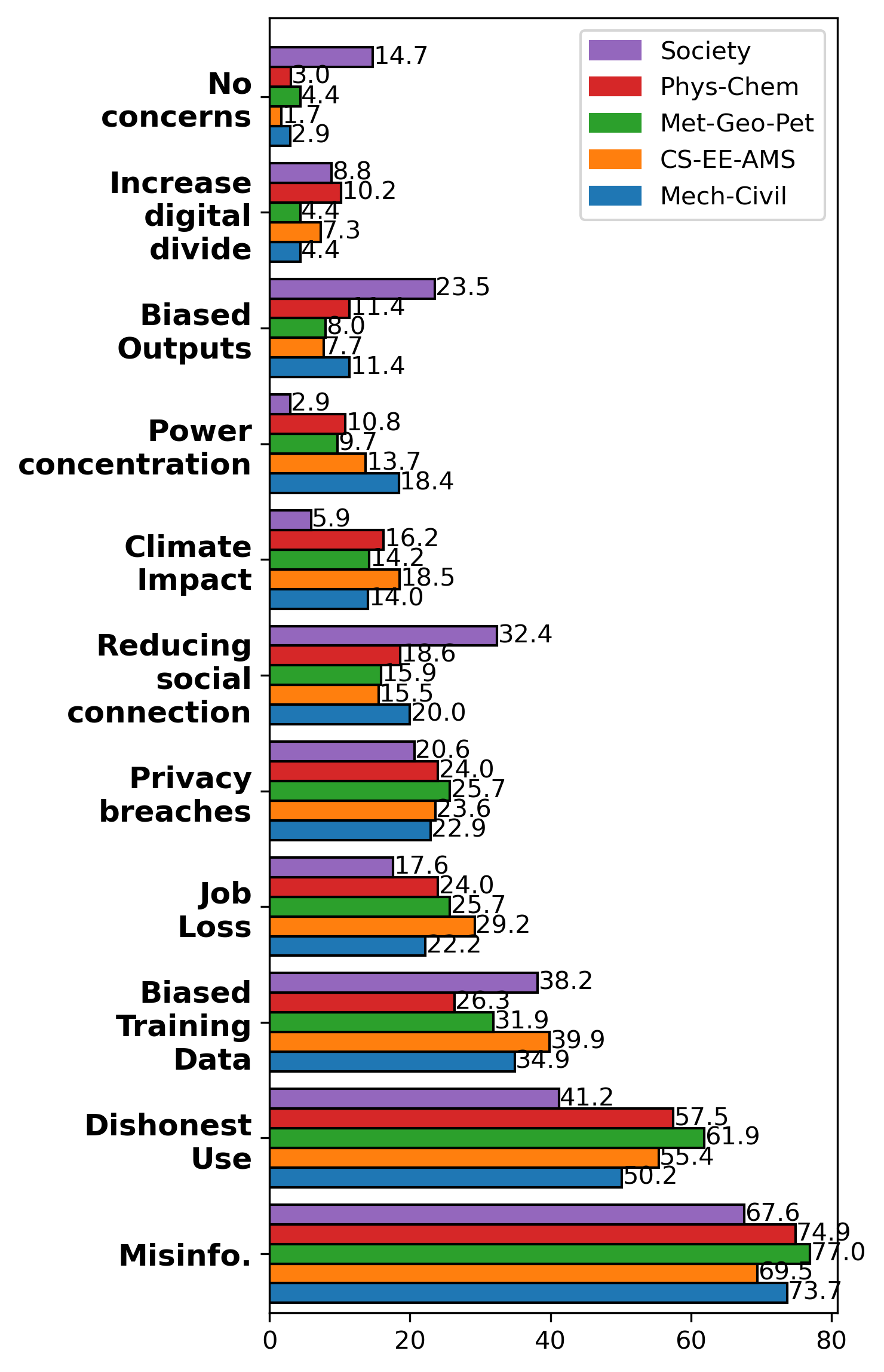}
    \caption{\textit{Students' Top Concerns Regarding GenAI in 2024 ($n=862$). X-axis: Frequencies normalized by \# of respondents per department cluster.}}
    \label{fig:concern}
    \vspace{-19pt}
\end{wrapfigure}
}

\newcommand{\pdoom}{
\begin{figure}[t]
    \centering
    \begin{subfigure}[b]{0.49\textwidth}
        \centering
        \includegraphics[width=\textwidth]{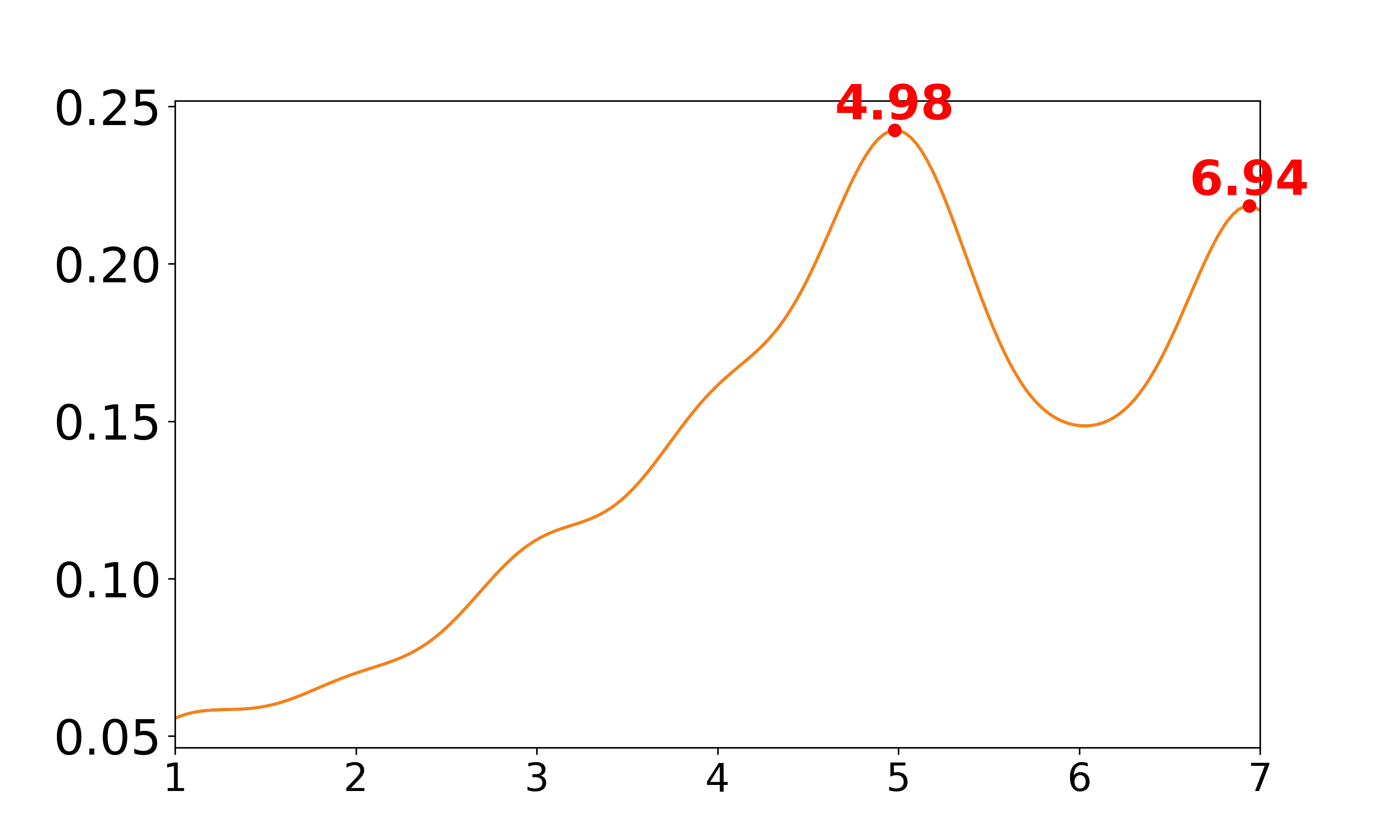} 
        \caption{\textit{``Importance of Considering P(doom)'' ($n=862$). X-axis: Likert ratings from (1) Unimportant to (7) Extremely important.}}
        \label{fig:imp}
    \end{subfigure}
    \hfill
    \begin{subfigure}[b]{0.49\textwidth}
        \centering
        \includegraphics[width=\textwidth]{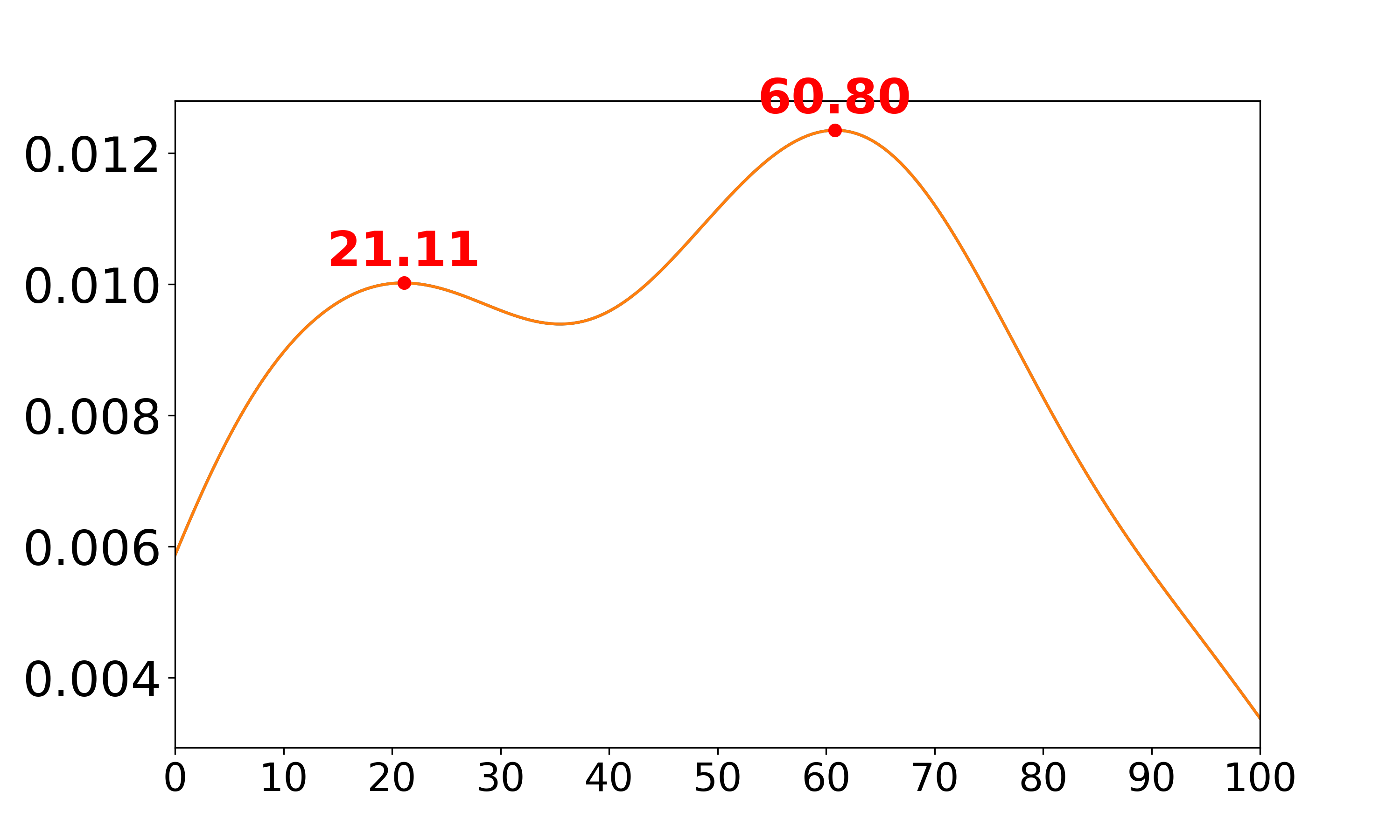} 
        \caption{`\textit{`Your Own Estimate of P(doom)'' ($n=664$). X-axis: P(doom) estimates from 0 (no chance) to 100 (certain catastrophe).}}
        \label{impor}
    \end{subfigure}
    \caption{\textit{KDE Plots of Student Perspectives on P(doom), a term referring to a person's estimated probability of AI causing catastrophic consequences for human society. Y-axes: Densities.}}
    \label{fig:p_doom}
\end{figure}
}

\begin{document}
\raggedright
\maketitle
\thispagestyle{empty}
\pagestyle{empty}

\section*{Abstract}

Generative Artificial Intelligence (GenAI) tools and models have the potential to re-shape educational needs, norms, practices, and policies in all sectors of engineering education. Empirical data, rather than anecdata and assumptions, on how engineering students have adopted GenAI is essential to developing a foundational understanding of students' GenAI-related behaviors and needs during academic training. This data will also help formulate effective responses to GenAI by both academic institutions and industrial employers. We collected two representative survey samples at the Colorado School of Mines, a small engineering-focused R-1 university in the USA, in May 2023 ($n_1=601$) and September 2024 ($n_2=862$) to address research questions related to (RQ1) how GenAI has been adopted by engineering students, including motivational and demographic factors contributing to GenAI use, (RQ2) students' ethical concerns about GenAI, and (RQ3) students' perceived benefits v.s. harms for themselves, science, and society. Analysis revealed a statistically significant rise in GenAI adoption rates from 2023 to 2024. Students predominantly leverage GenAI tools to deepen understanding, enhance work quality, and stay informed about emerging technologies. Although most students assess their own usage of GenAI as ethical and beneficial, they nonetheless expressed significant concerns regarding GenAI and its impacts on society. We collected student estimates of ``P(doom)'' – an informal AI safety term which expresses an individual’s estimated probability of catastrophic consequences stemming from AI – and discovered a bimodal distribution. Thus, we show that the student body at Mines is polarized with respect to future impacts of GenAI on the engineering workforce and society, despite being increasingly willing to explore GenAI over time. We discuss implications of these findings for future research and for integrating GenAI in engineering education.

\section*{Introduction}\label{sec:intro}
Recent advancements in Generative Artificial Intelligence (GenAI), esp. large language models (LLMs) like ChatGPT, have significantly impacted both industry and educational sectors~\cite{makridakis_forthcoming_2017, mohammed_role_2023}. These models, equipped with sophisticated algorithms and trained on vast datasets, can understand and generate human-like text~\citep{zhao_more_2023}, expanding their use from simple text prediction to composing essays, assisting with research, generating creative ideas, and even writing 
code. 
Progress in GenAI has redefined industry practices, impacting areas like customer service, creative writing, and software development~\citep{singh_transforming_2024}. Additionally, these technologies have emerged as powerful tools in education by facilitating new learning methods, enabling personalized academic support, and enhancing instructional delivery~\citep{ayeni_ai_2024}.

Integrating GenAI into education, particularly in engineering programs, presents new opportunities to enrich teaching and learning processes~\citep{yelamarthi_exploring_2024}. The engineering discipline, known for its emphasis on complex problem-solving, design innovation, technology adoption, and deep theoretical analysis, may benefit from GenAI capabilities such as content suggestions, AI-assisted simulations, real-time feedback for problem-solving, and interactive learning experiences. Consequently, students can access on-demand tutoring and resources that complement traditional classroom instruction. However, GenAI also introduces challenges, such as the need for ethical guidelines, the risk of academic dependency, and the importance of fostering critical thinking skills within AI-assisted learning environments~\citep{c_f_ho_communicating_2024,young_chemistry_2024}.

Understanding how engineering students adopt GenAI tools is critical for developing effective strategies for their integration into education. Preliminary studies such as~\citep{smith_early_2024,rogers_attitudes_2024,prather_robots_2023} have found that students in computing fields generally hold positive views toward GenAI tools for tasks like writing, coding, and learning. These investigations were conducted during early stages of GenAI adoption, with limited guidance on how to use the tools efficiently. 
A comparative perspective across diverse engineering disciplines can reveal evolving patterns of usage, highlight changes in perceptions, and provide insights into the broader impacts of GenAI on education and professional development. This approach is particularly valuable for informing institutional policies and ensuring that they are inclusive, equitable, and aligned with the diverse needs of engineering students. 
Therefore, this study poses three research questions: \textbf{(RQ1)} \textit{How have engineering students adopted GenAI, and what motivational and demographic factors contribute to its usage?}  
\textbf{(RQ2)} \textit{What are engineering students' ethical concerns related to the adoption of GenAI-based tools in their education and future careers?}
\textbf{(RQ3)} \textit{What are students' perceptions of benefits v.s. harms of GenAI for themselves, science, and society?}

To address these RQs, we conducted two separate surveys of all students at the Colorado School of Mines (henceforth referred to as ``Mines''), a small ($<10$k students) engineering-focused R-1\footnote{\textit{According to the \href{https://carnegieclassifications.acenet.edu/}{Carnegie Classification of Institutions of Higher Education}, R-1 indicates universities that offer doctoral degrees and have ``very high research activity.''}} university in the USA in May 2023 ($n_1=601$) and Fall 2024 ($n_2=862$). This paper reports on the findings, demonstrating statistically significant results related to increasing frequency of use of GenAI by engineering students over time, and other trends related to use cases, student motivations, ethical concerns, and perceived benefits \textit{v.s.} harms of GenAI.

\section*{Related Work}
This section describes the emerging literature on GenAI in educational practices and student experiences across engineering by summarizing what is known about its current adoption, use cases and pedagogical frameworks, and student/educator perspectives.

\paragraph{Current Adoption, Use Cases, and Pedagogical Strategies for Generative AI} \phantom{Previous} 
Previous research has evaluated how students have begun to adopt GenAI. Students have used GenAI to enhance problem-solving skills, deepen understanding of core subjects, assist with data analysis, foster interdisciplinary learning~\citep{harris_generative_2024}, idea generation, writing, coding, learning, and improving task efficiency~\citep{subramanian_artificial_2024,smith_early_2024}. Some research focused on LLM performance for common engineering problems such as solving simultaneous equations~\citep{riley_solving_2024}, completing programming problems of various difficulties~\citep{jayachandran_analysis_2023}, and completing mechanical engineering assignments~\citep{wilson_motivating_2024}. LLMs can perform well at many of these tasks, although they sometimes propagate incorrect solutions, misinformation, or out-of-scope information~\citep{jayachandran_analysis_2023,wilson_motivating_2024}. Overall, these studies highlight how students can readily utilize LLMs for a variety of use cases in academic work. 

Since GenAI is readily amenable to student use cases, prior work offers initial strategies for integrating GenAI in engineering education. Students believe they would benefit from being taught how to effectively use GenAI and being given clear instructions on how they are allowed to use these tools in academic settings~\citep{saiedian_leveraging_2024,smith_early_2024}. Proposed methods of integrating GenAI include re-designing classroom activities/assignments and restructuring grading schemes to focus more on the learning process rather than final results (which might be generated by GenAI tools and then copy/pasted) and using GenAI instead to provide tailored help to students~\citep{wilson_motivating_2024}. Studies have also explored how GenAI tools can change generative design processes in engineering~\citep{clay_board_2024}. 

Overall, prior studies suggest that the thoughtful integration of GenAI tools holds potential for revolutionizing personalized learning and assessment by providing powerful new student resources. However, these studies have largely been conducted in small subsets of engineering disciplines~\citep{subramanian_artificial_2024,peuker_evaluation_2024,smith_early_2024}. There is limited prior work that systematically evaluates how the adoption of GenAI has evolved over time across the population of an entire engineering university, leading to \textbf{(RQ1)} \textit{How have engineering students adopted GenAI, and what motivational and demographic factors contribute to its usage?} 

\paragraph{Student and Educator Perceptions of Generative AI}
Research indicates that students tend to have positive perceptions of GenAI tools like ChatGPT. Students tend to view GenAI as inevitable~\citep{saiedian_leveraging_2024}, and also feel that these tools can substantially improve their quality of work and overall learning~\citep{peuker_evaluation_2024}. Yet research on chemistry students’ AI literacy also suggests that students' ability to critically assess AI-generated content varies, raising concerns about over-reliance on potentially incorrect information~\citep{young_chemistry_2024}. Consequently, educators have a range of opinions regarding students' use of GenAI tools. One study found that some educators agree with the predominant student perspective that GenAI has the potential to improve education while others are skeptical, citing the potential of GenAI to facilitate academic dishonesty~\citep{harper_faculty_2024,osunbunmi_board_2024}. Yet educators have not adopted GenAI at the same rate as students, leading to tensions and misunderstandings between students and teachers~\citep{prather_robots_2023}. Prior work has largely focused on whether students enjoy using GenAI tools but does not explore ethical concerns regarding their use, nor perspectives on broader impacts of GenAI across society, thus motivating our final RQs. \textbf{(RQ2)} \textit{What are engineering students' ethical concerns related to the adoption of GenAI-based tools in their education and future careers?} \textbf{(RQ3)} \textit{What are students' perceptions of benefits v.s. harms of GenAI for themselves, science, and society?} Addressing these questions will provide empirical insights that could demystify student perspectives for educators, while capturing an important historical benchmark of engineering students' attitudes toward GenAI during its early introduction to society.

\section*{Methods}\label{sec:methods}
This section details our survey designs, samples, recruitment, and analysis. Our study was reviewed and deemed exempt by the Human Subjects Research board at Mines.

\paragraph{Survey Design}
We conducted our first survey in May 2023, as soon as institutionally possible following the release of ChatGPT in November 2022. The first survey was a preliminary and exploratory instrument that helped us to qualitatively discern use case categories and concerns of interest. It used primarily open-ended free response questions about students' use cases and perspectives on GenAI, which we analyzed using Directed Content Analysis~\citep{hsieh_three_2005}.\footnote{\textit{Details of our codebook development are reported in~\cite{smith_early_2024}, a preliminary paper that reported on only the Computer Science subset of the full 2023 survey data.}} It also included Likert rating questions on frequency of GenAI use and estimated benefits v.s. harms of GenAI to a student's scientific field. 
To design the second survey, we began by duplicating the first survey's structure and questions. We retained \textit{exact} question phrasings for Likert rating questions on student's frequency of use and perceived benefits of GenAI to their engineering field so that we could report on direct comparisons  over time. However, we replaced the original free response questions on use cases with new multiple-choice questions with discrete answer choices derived from the results and codebooks of the first survey. This approach ensured that the 2024 answer responses were empirically derived and highly relevant to our student population, while offering a finer level of granularity and quantifiability across the population than in 2023. Because the surveys used different and non-comparable question-asking techniques to collect information about student use cases, motivations, ethical concerns, etc., most of our results focus on the 2024 data. 
We piloted and refined both survey versions extensively with members of our research team as well as our university's center for educational innovation in order to identify bugs, potentially confusing questions, and conceptual gaps in question coverage.

In 2024, final survey question blocks included: consent and eligibility; 
ratings of frequency of use of a variety of GenAI-based tools for classes or professional efforts; LLM use cases; motivations for LLM usage; concerns with GenAI; whether GenAI should be allowed in courses; perceived benefits \textit{v.s.} harms of LLM usage; the importance of considering P(doom); and estimates of P(doom). As defined by Wikipedia~\citep{wikipedia_pdoom_2024}, P(doom) is an informal AI safety term expressing an individual’s estimated probability of catastrophic consequences for human society stemming from AI.\footnote{\textit{To our knowledge, no prior academic research has defined or measured P(doom). We aimed to define this term in a way that is accessible and understandable for all participants. Wikipedia provides a widely recognized, straightforward explanation suitable for this purpose.}} We intentionally opted not to use the term ``cheating'' as a use case because participants may be reluctant to self-identify as engaging in academic dishonesty. Instead, we included the use case option, ``generating solutions without first solving the problem,'' which can be interpreted as an indicator of potential misuse or shortcuts in problem-solving. We also asked how ethical students felt their use of LLMs was, as  another potential proxy for cheating.


The 2023 survey collected limited demographic information (undergrad, masters, or doctoral status; \# years enrolled, and department); the 2024 survey also collected 
 socioeconomic status (SES, MacArthur Scale of Subjective Social Status~\citep{adler_macarthur_2000}), race, gender, and disability status. 
 To protect anonymity, neither survey collected name, email, or other personally identifiable information. To support replications, verbatim survey questions are available as supplemental materials at \href{https://bit.ly/genAI-survey-text}{bit.ly/genAI-survey-text}. 
 Both surveys were built in QuestionPro software.

\paragraph{Sample Selection}
The target population we seek to describe is all students at the Colorado School of Mines, a small engineering-focused university in the USA. This population provides a valuable, in-depth glimpse into how GenAI has been adopted across diverse engineering and STEM disciplines at one university site. 
In Fall 2023, total reported enrollments included 5,825 undergraduate and 1,756 graduate students, for a total of 7,608 students~\citep{colorado_school_of_mines_enrollment_2024}. 
For this target population size ($N=7608$), we calculated that we required a sample size of at least ($n=507$) to achieve a 98\% confidence level ($\alpha=0.02$) with a 5\% margin of error ($E=0.05$).


\paragraph{Recruitment}
The Office of the Provost permitted us to send recruitment emails to two list-servs, one each for all current undergraduate and graduate students when the surveys opened, with a follow-up reminder the day before they closed. We ran the first survey May 2-19, 2023, and the second, September 4-13, 2024. 
After completing the survey, participants were routed to a separate form where they could opt-in to enter their email in a drawing for one of four \$25 gift cards. Table~\ref{table:respondents} summarizes our participants. 
Both samples exceeded the calculated sample size and can be considered representative of the student population with 98\% confidence.

\paragraph{Survey Analysis}\label{sec:DCA}
Data were exported from QuestionPro into a CSV file. 
To ensure data quality, we first performed cleaning steps by removing: incomplete responses, responses with excessively rapid completion times, responses that demonstrated column-filling patterns (i.e. a participant marked all responses in the same column very quickly, indicating that they did not read the response choices), and responses that contained nonsensical free response data. 
The 2024 survey focused on multiple-choice questions with discrete answer choices, however  questions on use cases, motivations, and concerns also included an option to write in an ``other'' response. We reviewed all written-in responses and re-labeled them as: irrelevant to the question (and ignored them); similar to a predefined answer choice (and counted it toward that option); or as a new concept  (and created a new label to expand the dataset's descriptive richness). Finally, since there are numerous departments of differing sizes at Mines, we clustered responses into five groups of departments based on our assessment of shared academic focus and disciplinary overlap (Table~\ref{table:department_clusters}). These clusters improve the coherence and visualization of our reporting by organizing the data and providing insight into how related departments are adopting GenAI. Following data cleaning, we used standard \texttt{pandas} and \texttt{scikit-learn} Python packages to compute descriptive statistics and statistical tests. Throughout results, we indicate which \textit{optional} questions were not answered by all survey respondents, and if so, how many respondents provided a response. (For these questions, proportions are reported out of \textit{only} the students who provided answers.) Finally, to allow for proportional comparisons across department clusters in bar graphs, we normalize frequencies of responses by number of respondents per cluster.

\paragraph{Threats to Validity}
Standard survey limitations apply to this study, such as opt-in bias and the possibility of faulty recollection or inaccurate self-assessment. Additionally, students may have altered information about behaviors perceived as cheating; to counteract this, we used messaging encouraging honesty because participation was anonymous and no data could be shared beyond the research team. Because students took the survey anonymously, we could not access data about actual performance (e.g., test grades, accuracy or efficiency of completing tasks, etc.); future work should combine self-report measures with measures of actual performance. Additionally, we could not run both surveys at the same time of semester due to administrative constraints; the first survey was at the end of Sp23 and the second in the beginning of Fall24. Table~\ref{table:respondents} shows a jump in the proportion of first year students taking the second survey, which may slightly confound comparisons over time (but does not impact analyses of the 2024 data alone, which are the majority of analyses in the paper). Future work should continue with re-surveys at consistent times of year, or at the beginning of every semester if possible. Finally, our survey data are from only one university site and therefore we cannot generalize our results across all engineering education. Future work should survey other sites to understand broader adoption patterns.

\deptclusters
\participants

\section*{RQ1 Results: Adoption of GenAI Across Engineering Disciplines}\label{sec:results}

\paragraph{Frequency of Use of LLM-Chatbots}
To investigate the adoption of GenAI tools over time, survey data from 2023 ($n_1 = 601$) and 2024 ($n_2 = 862$) were analyzed to determine changes in usage frequencies of LLM-powered chatbots among students. In both surveys, all respondents received a required question (verbatim both years) to rate frequency of use of a variety of GenAI tools in relation to their academic pursuits as ``never,'' ``once or twice'' (i.e., irregular users), ``regularly (once or twice a week),'' ``all the time (daily, or more often)'' (i.e., superusers), or ``only for fun, never work.'' We report first on LLM-chatbots, and next other types of GenAI.

\adoptionrate

Our data suggest rising utilization of LLM-powered chatbots in students' academic routines. 
The proportion of students reporting never using LLM chatbots significantly decreased from 30.8\% in 2023 to  17.9\% in 2024 ($\beta=+5.76$, $p<0.001$). 
Conversely, while there was no significant change in the proportion of irregular users, the proportion of both regular users and superusers showed significant increases over the same period. Regular users increased from 22.46\% in 2023 to 32.3\% in 2024 ($\beta = -4.09$, $p<0.001$), while superusers rose from 9.3\% in 2023 to 12.8\% in 2024 ($\beta = -2.04$, $p<0.05$). Thus, these data suggest that by 2024, 45.1\% of students are regular or superusers of LLMs, with an additional 17.7\% who have used them irregularly. 

Figure~\ref{fig:frequencyusaecase} illustrates that the frequency of LLM-powered chatbot usage appears to vary across the different department clusters. 
Further analysis of the data revealed significant differences in LLM usage across department clusters ($\chi^2 = 36.07$, $p < 0.01$, $DF = 16$). Students in the CS-EE-AMS cluster were significantly more likely to use GenAI tools ``all the time'' ($Residual (R) = 3.2$, $p < 0.01$), while those in the Mech-Civil ($R = -2.6$, $p < 0.01$) and Phys-Chem clusters ($R = -1.6$, $p < 0.1$) were less likely to be superusers. Notably, the Phys-Chem cluster also had a higher proportion of students never using these tools ($R = 2.2$, $p < 0.05$). 

\paragraph{Use of Other GenAI Tools}
Our analysis indicates that most students have \textit{not} used other types of GenAI tools. Although the proportion of students who reported never using image generators decreased from 71.4\% in 2023 to 58.1\% in 2024 ($\beta = 5.185$, $p<0.001$), and irregular users of image generators increased significantly from 14.81\% to 23.09\% ($\beta = -3.917$, $p<0.001$), there were no significant changes in regular or superusers of image generators. These data suggest that of the students curious about image generators, most have only experimented with them once or twice. Most students (80.5\%) in 2024 have also never used other types of media generators (e.g. song, voice, video, etc.) Because of low utilization of non-LLM forms of GenAI, we exclude other GenAI tools from further analysis and focus the remainder of the paper on LLM-chatbots. 
\usecases
\paragraph{Student Use Cases for LLM-Chatbots}
Given that many students have adopted LLM-chatbots, we further explore the diverse purposes for which students leverage these tools across three key domains: learning, coding, and writing. Multiple-choice questions on use cases were presented only to irregular, regular, and superusers ($n = 645$)---i.e. students who had never used LLMs, or only for fun, did not receive the question. (Written-in responses yielded no new use cases.)

\textbf{Learning:} Figure~\ref{fig:learning} summarizes learning-related LLM use cases. A significant proportion of students use LLMs for enhancing their understanding of course materials by exploring unfamiliar concepts (52.4\%), explaining challenging topics (62.3\%), and breaking down complex problems into manageable parts (47.6\%). Many students also leverage these tools as alternatives to traditional learning tools. Specifically, they use LLM-powered chatbots to replace search engines (55.7\%), non-interactive resources (52.1\%) (e.g., textbook, documentation), or interactive learning resources (27.4\%) (e.g., emailing the professor, office hours, or course discussion forum). Students also utilize LLMs for validating their own original solutions (47.9\%) or--more problematically, but least commonly--generating solutions to problems without first solving them (20.8\%). 49.1\% also rely on LLMs for summarizing textual content, making complex reading materials more accessible. These findings highlight the diverse ways in which LLM-powered chatbots are being used to support students' learning and academic work, with a strong focus on conceptual understanding and replacing traditional educational resources.

\textbf{Coding:} Figure~\ref{fig:coding} depicts participants' coding use cases. Interestingly, students' preferred coding use cases are understanding the behavior of code (44\%)  and debugging or troubleshooting code (48.4\%) rather than drafting code (26.5\%), accelerating the development of code (28.5\%), or converting code to different languages (15.5\%). This suggests that students often see significant value in using LLM-powered chatbots to assist with understanding program functionality and error identification---tasks that are often time-consuming and challenging---rather than completing coding tasks for them. 
Nonetheless, these findings highlight the dual role of LLM-chatbots in supporting problem-solving and enhancing productivity.

\textbf{Writing:} Figure~\ref{fig:writing} shows participants' writing use cases for LLM-chatbots, which underscore the role of chatbots in fostering creativity and providing a starting point for research or project development. For example, 59.5\% of students use LLMs for brainstorming new ideas and 23.3\% use them to collect resources to support their work (e.g., using LLMs to create lists of examples or references, and then looking them up). Students also use LLM-chatbots for outlining documents (33.5\%), generating initial drafts (20\%), or improving and refining drafts that they had created (46.7\%) to improve clarity, coherence, concision, or grammar. Thus, students value these tools as practical aids for streamlining both early and later stages of the writing process.



\textbf{Motivational factors} To explore factors driving students' use of LLMs, the survey included an optional multiple-choice question asking participants to select up to three most important motivations influencing their decision to use LLMs. 142 participants responded, providing an average of 2.66 motivations per student. (Written-in responses yielded no new motivations.)

Figure~\ref{fig:motivation} provides insights into underlying motivations for student engagement with GenAI tools. We observe that students more often indicated positive intent to incorporate these tools to benefit their knowledge and academic routines, rather than less savory motivations related to grades, overwhelm, etc. The most commonly reported motivations were to enhance understanding and gain deeper insights into subjects ($n=82, 57.7\%$),  
to improve the quality of academic or professional work $(n=57, 40.1\%)$, 
and curiosity to  experiment with cutting-edge AI tools ($n=47, 33.1\%$). These findings underline the multifaceted appeal of LLM-powered chatbots in engineering education, driven by a mix of academic goals and technological curiosity. Nonetheless, smaller proportions of students also report motivations related to not seeing the value in assigned work ($n=10, 7.1\%$), being too overwhelmed ($n=23, 3.6\%$), earning good grades ($n=26, 4\%$), or increasing their free time for their personal life ($n=28, 4.3\%$).
\motivations 

\textbf{Demographic Factors}
We did not identify any significant correlations between frequency of GenAI usage and the number of years enrolled, gender, or disability status. However, the data demonstrated that graduate students exhibit higher GenAI usage compared to undergraduates, with significantly more superuser graduate students ($Residual (R) =4.5$, $p<0.01$) and fewer irregular users ($R = -1.7$, $p < 0.1$). Undergraduates were significantly underrepresented in the superuser category ($R = -2.6$, $p < 0.01$). Furthermore, the analysis revealed a significant association between socioeconomic status (SES) and frequency of chatbot usage ($\chi^2 = 22.8$, $p < 0.01$, $DF = 8$). Low SES individuals were significantly less likely to be irregular users ($R = -2.13$, $p < 0.05$) and more likely to be superusers ($R = 3.07$, $p < 0.01$). In contrast, usage patterns of medium and high SES individuals generally aligned with expected frequencies. Additionally, racial/ethnic identity may influence adoption rates of GenAI. We found that Asian participants were overrepresented in the superuser category ($R = 2.6$, $p < 0.01$) and underrepresented in the never category ($R = -2.1$, $p < 0.05$). Conversely, white participants were less likely to be superusers ($R = -2.0$, $p < 0.05$), while usage patterns of participants from other racial/ethnic backgrounds were generally consistent with expected frequencies.

\studentethics

\paragraph{Use of Free v.s. Paid LLM-Chatbots} 
The 2024 survey (optionally) asked whether students had ever paid for LLM-chatbots (e.g., ChatGPT-4). Of the 140 students who provided an answer, most are free users who have never paid for these services, however 9 ($6.4\%$) reported having paid in the past but discontinued their subscriptions, while only 10 ($7.1\%$) are currently paying subscribers. (Notably, among the current paying users, 7 belong to the CS-EE-AMS cluster, hinting at a stronger inclination towards premium features within this group.)

We did not identify a significant correlation between SES and subscription status. However, we found that frequency of GenAI usage significantly correlates with subscription status ($\chi^2=38.3752$, $p<0.01$, $DF=8$). Students who are currently paying for GenAI subscriptions are not irregular users ($R = -1.7$, $p<0.1$) but are significantly more likely to be superusers ($R =5$, $p<0.01$). Conversely, students who reported never paying for GenAI subscriptions were less likely to be superusers ($R = -1.8$, $p<0.1$). Usage patterns of students who had previously subscribed but no longer maintained a subscription were consistent with expected frequencies.

\section*{RQ2 Results: Ethical Concerns about GenAI}\label{sec:rq2}

\paragraph{Student Perceptions of Ethical Use}
We explored students' perceptions of how ethical they perceived their \textit{own} use of LLM-powered chatbots to be on a scale of 1 (completely unethical) to 7 (completely ethical) (Figure~\ref{fig:ethical}). Among the 651 students who answered the question, the mode of responses was 7, and second most common selection 6. This finding demonstrates a clear tendency among students to view their use of LLM-powered chatbots as ethically acceptable, reinforcing their positive attitude toward integrating such tools into their academic and professional activities.


\paragraph{Student Concerns with GenAI}
Although most students believe their own use of LLMs to be ethical, the integration of GenAI in academic settings nonetheless raises important ethical and societal concerns. The survey included a required question asking participants to select up to three top concerns with GenAI. 862 participants responded, with an average of 2.78 concerns per student (Figure~\ref{fig:concern}). We identified several additional concerns in written-in ``other'' responses including: the ability of GenAI to replace real learning ($n=29$), replacing human creativity ($n=8$), and incentivizing laziness ($n=2$).
\concerns 
Misinformation emerged as the most prominent concern ($n = 630, 73\%$), indicating that students are highly aware of the potential for AI generated content to spread inaccurate or misleading information. This concern is particularly relevant in light of recent incidents where AI tools have produced content with factual inaccuracies, highlighting the need for stronger verification processes and responsible AI use in educational contexts. The second highest concern ($n=468, 54.2\%$) was dishonest use of AI. Many students are apprehensive that some may use GenAI to complete assignments or projects dishonestly, undermining academic integrity. This concern raises questions about the value of authentic effort in education and emphasizes the need for policies and practices that encourage ethical use of AI tools while discouraging misuse. Another significant concern ($n=296, 34.3\%$) related to biased training data, such as gender or racial biases, underscoring a need for transparency in AI systems and highlighting a need to address bias during AI development to ensure equitable outcomes.

\phantom{phantom}

\paragraph{Integration of GenAI in Classes}
One required multiple-choice question asked students about how GenAI should be integrated into their coursework. The vast majority of students ($n= 792$, $91.9\%$) expressed support for the use of GenAI in academic settings. Of these, 511 ($59.3\%$) felt that each instructor should determine the most appropriate GenAI policy for their class. On the other hand, others favored a ``blanket'' policy applicable to all courses on campus: 83 ($9.6\%$) wanted unrestricted use of GenAI in all classes, whereas 198 ($23\%$) preferred that GenAI should be allowed in a restricted way, such that instructors must specify which exact uses or applications of GenAI are and are not permissible in each individual class. Finally, despite broad support for GenAI, a smaller group of students, $(n=71, 8.2\%)$ opposed the use of GenAI in coursework altogether, citing concerns over its ethical implications and potential for misuse. 

\section*{RQ3 Results: Perceived Benefits v.s. Harms of GenAI}
In 2023, all survey respondents were asked to rate perceived benefits v.s. harms of GenAI to ``your scientific field or major'' on a scale of 1 (extremely harmful) to 10 (extremely beneficial). In 2024, this question was repeated verbatim, however we \textit{additionally} requested ratings for benefits v.s. harms to: your learning during higher education; your efficiency at completing tasks; the quality of tasks you complete; your job placement after graduating; your intended job itself after graduating; society at large. We use Kernel Density Estimate (KDE) plots to visualize the probability densities of the ratings, providing a smoothed representation of the distributions. Peaks indicate where responses are concentrated, while area under the curve represents the overall distribution of responses, summing to 1. (Note: Y-axes are probability densities, not proportions of responses.)

\paragraph{Decline in Perceived Benefits to Scientific Field}
Our analysis revealed a significant decrease in students' rating of perceived benefit to their scientific field or major from 2023 to 2024 ($\beta = 12.31$, $p<0.01$) (Figure~\ref{fig:fieldbenefits}). Students in 2023 perceived GenAI as more beneficial to their field (median = 7) compared to students in 2024 (median = 6).

\scientificfield
\ratings

\paragraph{Neutral to Positive Leanings in Perceived Benefits to Self and Society in 2024}
Although job loss or displacement was a ``top concern'' for some students (we note that it was the fourth most common selection in Figure~\ref{fig:concern}), students in 2024 overall trended toward expressing a neutral perception of the current impact of GenAI on their job placement after graduation, intended jobs themselves, and society at large: Figures~\ref{fig:jobplacement},~\ref{fig:intendedjob}, and~\ref{fig:societyatlarge} depict KDE plots of rating distributions that all peak around 5. Alternatively, students exhibited somewhat more positive inclinations when considering GenAI's impact on their learning, task efficiency, and task quality (Figures~\ref{fig:yourlearning},~\ref{fig:yourefficiency}, and~\ref{fig:taskquality}). Taken together, these findings suggest that students tend to feel neutral about the current impacts of GenAI on their careers and society, whereas they are more optimistic about how GenAI is now enhancing their own learning and productivity. We note that for some students, there may exist cognitive dissonance between AI \textit{today}, versus AI of the \textit{future}, as our results on P(doom) below indicate.

\paragraph{Higher Frequency of LLM Usage Correlates with Higher Benefit Ratings}
Interestingly, we observed that a higher frequency of LLM usage is associated with higher benefit ratings in the 2024 data. We report in detail on benefits to one's major or scientific field ($\chi^2 = 208.7$, $p<0.01$, $DF = 36$), however we note that the trends are similar across all types of benefits v.s. harms assessed. For example, students who reported never using LLMs tended to provide low benefit ratings ($Residual (R) = 5.0, 3.2$, and $3.1$ for ratings $1, 2$, and $3$, respectively, $p<0.01$), and not to provide high benefit ratings ($R = -2.5, p<0.01$ and $-2.4, p<0.05$ for ratings 8 and 10, respectively). Conversely, regular users exhibited an opposite trend, with significant positive residuals for benefit ratings 6 and 9 ($R = 2.7, p<0.01$ and $2.2, p<0.05$, respectively) and significant negative residuals for ratings 1, 3, and 4 ($R = -3.0, p<0.01, -2.2, p<0.05$, and $-2.4, p<0.05$, respectively). Similarly, superusers had significant positive residuals for high benefit ratings 8, 9, and 10 ($R = 3.3, p<0.01, 2.1, p<0.05$, and $5.5, p<0.01$ respectively) and significant negative residuals for low benefit ratings 1 and 3 ($R = -2.1$ and $-2.4, p <0.05$ respectively). Overall, these findings suggest that the more frequently students use LLMs, the more beneficial they perceive them to be.

\pdoom

\paragraph{Student Perceptions of P(doom)}
To gauge students' imagined impacts of AI on society in the future, we explored their views on P(doom). We provided verbatim text from Wikipedia~\cite{wikipedia_pdoom_2024} (along with a citation to Wikipedia) that ``P(doom) refers to the probability of catastrophic outcomes (or ``doom'') as a result of artificial intelligence.'' We first asked a required question about how important respondents felt it was ``to consider P(doom) when using or developing AI?'' on a scale of 1 (unimportant) to 7 (extremely important). Next was an optional question to ``share your own estimate for P(doom) by entering a number between 0 (no chance) and 100 (certain catastrophe).'' 

Figure~\ref{fig:imp} shows that most students felt it was moderately or highly important to consider P(doom); 531 students provided a rating $\geq5$, while 331 gave a rating $<5$. While the overall mean was 4.8 ($SD = 1.8$), the KDE plot shows two distinct peaks at 5 and 7, confirmed as a bimodal distribution by a kurtosis test ($Kurtosis = -5.8$, $p < 0.05$). This distribution highlights the general recognition among students of the importance of evaluating risks when using or developing AI tools.

Moreover, Figure~\ref{impor} demonstrates a striking split in the student population in estimates of P(doom). Among 664 responses, the mean estimate was 46.7 ($SD = 28.0$). However, the KDE plot highlights two prominent clusters, confirmed by a kurtosis test as another bimodal distribution ($Kurtosis = -12.9, p < 0.05$), with one peak around 21.1, representing students who are relatively less concerned, and another at 60.8, reflecting higher concern. Separating the students into two subgroups delineated by the inflection point between these two peaks at 40, 44\% of the population ($n=290$) estimated low P(doom) while the other 56\% ($n=373$) estimated high P(doom). This finding demonstrates a substantially polarized student body, with many acknowledging substantial risks associated with AI while others remain more optimistic about its safety.

\paragraph{Demographic Associations with P(doom)}
Our analysis further examined whether demographic factors or academic backgrounds influenced students' perceptions of P(doom) regarding AI technology. Results revealed a significant difference in socioeconomic status (SES) between the two groups—those who assigned low versus high ratings to P(doom) ($\beta = 6.6$, $p<0.01$). Students with higher SES were more likely to assign lower P(doom) ratings ($Coef = -0.14$, $p<0.01$). However, no significant correlations were found between P(doom) and other factors such as department cluster, gender, or degree enrollment.

These findings, combined with students' generally neutral-to-positive perceptions of GenAI's practical applications, suggest that while students appreciate the utility of GenAI in their academic and professional endeavors, a notable proportion remains cautious about its broader societal and existential implications. The diversity in P(doom) ratings underscores the complex challenges of integrating GenAI into education and society responsibly.

\section*{Discussion}
Our findings provide a comprehensive view of engineering students' perceptions, motivations, and concerns regarding the integration of Generative AI (GenAI) in their academic and professional environments. 
By exploring students’ adoption patterns, motivations, ethical concerns, and perspectives on risks associated with GenAI, we uncover interesting insights that inform the responsible use of GenAI in higher education and establish directions for future research.

\textbf{Themes in Student Usage of GenAI}
Our data demonstrate that use of GenAI at the Colorado School of Mines is rising year-over-year, with students primarily using free versions of models. In 2024, 12.8\% of engineering students report using LLMs nearly everyday, while another 32.3\% use them a few times per week. Educators should be aware that around half of their students are using LLMs often, while the other half are either not using them at all, or only irregularly.

Although around a quarter of students admitted to using GenAI to generate solutions before solving a problem, Figure~\ref{fig:learning} shows that many \textit{more} students report using GenAI to deepen learning, while Figure~\ref{fig:motivation} demonstrates that positive motivations  for LLM use (e.g., increasing depth of knowledge and quality of work) are more common than more negative motivations (e.g., not seeing value in assignments or being too overwhelmed). An oft-repeated concern regarding GenAI is that its ability to automate creative tasks like coding and writing will detract from the pedagogical value of those tasks.
The analysis of student use cases in Figure~\ref{fig:usecases} reveals that more students are using GenAI for editing and critique tasks, rather than creative tasks.
For example, Figure~\ref{fig:coding} shows that students use GenAI for debugging and explaining code (editing/critique) rather than drafting code (creation).
Similarly, Figure~\ref{fig:writing} indicates that students use GenAI for brainstorming, outlining, and improving drafts (editing/critique) more than for generating an initial draft (creation).
On its face, these are promising trends; most students appear to be creating answers on their own, and then seeking assistance and feedback from automated tools.

An alternative perspective is that the focus on creation over editing and critique may actually be the wrong direction.
For example, debugging and code tracing are well-established elements of computer science pedagogy.
These tasks encourage students to develop their own models of what a program is doing, and programming bugs can illustrate fundamental misconceptions which can then be corrected.
Replacing the debugging, editing, and critiquing tasks with GenAI may reduce students' ability to address and rectify their own misunderstandings and mistakes.

Nearly 92\% of surveyed students believe that GenAI should be integrated into their coursework. 
Further research is required to examine if the current trends towards using GenAI for editing and feedback have pedagogical merit, or if reactionary educational policies emphasizing ``doing your own work'' have it backwards.
Concurrent research is already examining the efficacy of having students critique GenAI output~\citep{steele_gpt_2023,griffin_learning_2016,mccauley_debugging_2008}; our findings here encourage future investigation in this area.

\textbf{Leveraging Student Concerns for Critical Engagement}
Figure~\ref{fig:concern} reveals which GenAI concerns are more or less important to students.
Areas of high concern such as misinformation, plagiarism, and biased training data can be effective initial topics for students because more students can build on personal experiences and concerns.
Activities and assignments can avoid the work of ``motivating'' the concern, and focus on the impact and ways of mitigating harm. Figure~\ref{fig:concern} also reveals which topics are \emph{not} of immediate concern to most students, including discriminatory outputs, concentration of power, climate impact, privacy breaches, and job displacement. 
These topics may be less effective at initially engaging students to think about the impact of GenAI, and may be more suitable for more advanced discussions. 
Determining why these are less relevant to students is a potential area of future work.

\textbf{Dissonance Between Current GenAI Optimism v.s. Future Impacts} 
Our investigation into students’ perspectives on P(doom) introduces a previously underexplored dimension. Scholars have critically argued that AI hype can drive premature adoption of AI despite known inequities and harms, and that a focus on longterm risks rather than immediate impacts can fuel the fire to invest more heavily in AI~\citep{gebru_tescreal_2024}. For example, a highly publicized survey from 2022 demonstrated that on average, AI experts estimate a 5\% chance of AI causing human extinction~\citep{stein-perlman_2022_2022}. In our survey, we wanted to understand student perspectives on risks of GenAI. We softened the language away from human extinction to use the popular term P(doom) which more broadly encompasses ``catastrophic consequences for humanity.''~\cite{wikipedia_pdoom_2024} 
Most students considered it moderately or extremely important to account for P(doom) when using or developing AI tools. Although our methods and data cannot address the causes of this, we posit that the complex zeitgeist of \textit{both} AI hype and angst may be influencing students' choices and perspectives. 

Students were also asked to provide estimates of P(doom). Compared to the 5\% probability of human extinction from AI experts, we observed a \textit{much} higher average estimate from students of 46.7\% probability of catastrophic consequences, as well as a striking bimodal distribution of responses with peaks at 21.1 (lower concern) and 60.8 (higher concern). Although students tended toward neutral or positive ratings of \textit{current} perceived benefits to themselves, science, and society (Figures~\ref{fig:fieldbenefits},~\ref{fig:perception1}), their estimates of P(doom) illustrate serious concern for the future. We interpret these results as suggestive of cognitive dissonance. Although students are adopting GenAI more and more over time, and are typically trying to use it for benefit rather than harm, they are also rating slightly \textit{less} benefit to science in 2024 relative to 2023, and are demonstrating alarmingly high concern about the long term impacts of GenAI. Educators should be aware that their students likely have differing rates of GenAI adoption, as well as highly polarized opinions about GenAI and its future possible benefits and harms. This divide suggests rich opportunities for classroom activities or assignments that can cultivate skills in critical thinking, risk assessment, ethical decision-making, and productive discourse about AI in light of such differing perspectives. 


\section*{Conclusion}
While previous work has studied applications and limitations of GenAI tools for engineering tasks~\citep{riley_solving_2024,jayachandran_analysis_2023,wilson_motivating_2024}, this study provides valuable insights into the perceptions, motivations, and ethical concerns of engineering students regarding GenAI. While students are generally optimistic about the practical benefits of GenAI for their academic and professional work, there is significant awareness of ethical and societal risks associated with AI. These results corroborate previous findings regarding ethical concerns over GenAI~\citep{harper_faculty_2024,osunbunmi_board_2024}. Our work reveals students' top concerns around misinformation, dishonest use, and bias, highlighting the need for responsible GenAI integration into education. By developing clear guidelines, fostering critical thinking, and offering students opportunities to engage with the ethical challenges of AI, institutions can help students navigate the complexities of GenAI in a responsible and informed manner.

\section*{Acknowledgments}
First, thanks to all the students who took our surveys--we genuinely appreciated the opportunity to hear from you! We also thank students Kylee Shiekh, Michael Ivanitskiy, Spencer Wood, and Grace Ary for contributions to preliminary analyses and drafts of this paper. We also thank Dr. Iris Bahar and the Computer Science Department at the Colorado School of Mines for providing funding support for this work, as well as to the Trefny Innovation Instruction Center, the Provost's office, and Academic Affairs for supporting and permitting repeated surveys of the student body.

\vspace{4\baselineskip}\vspace{-\parskip} 
\footnotesize 
\bibliographystyle{IEEEtran}

\bibliography{generative-ai}

\end{document}